\documentstyle[preprint,revtex]{aps}
\newcommand{\KSM}{ $ K^0 $ - $ \bar{K}^0 $ system }
\newcommand{\DSDQ}{ \Delta S = \Delta Q }
\newcommand{\DSMDQ}{ \Delta S = - \Delta Q }
\newcommand{\wt}{{\it \Gamma}}
\newcommand{\ws}{{\it \Gamma_S}}
\newcommand{\wl}{{\it \Gamma_L}}
\newcommand{\taus}{{\it \tau_S}}
\newcommand{\dt}{{\it \Delta t}}
\newcommand{\adt}{\left| \dt \right|}
\newcommand{\dm}{{\it \Delta m}}
\newcommand{\Rp}{{\rm Re}\;}
\newcommand{\Ip}{{\rm Im}\;}
\newcommand{\BD}{\bar{{\it \Delta}_l}}
\newcommand{\KT}{ \left| K (t) \right> }
\newcommand{\KBT}{ \left| \bar{K} (t) \right> }
\newcommand{\KS}{ \left| K_S \right> }
\newcommand{\KL}{ \left| K_L \right> }
\newcommand{\KO}{ \left| K^0 \right> }
\newcommand{\KB}{ \left| \bar{K}^0 \right> }
\newcommand{\PL}{ P_{l^-} }
\newcommand{\PA}{ P_{l^+} }
\newcommand{\BL}{ \bar{P}_{l^-} }
\newcommand{\BA}{ \bar{P}_{l^+} }
\newcommand{\pst}{ \left| \psi (t) \right> }

\renewcommand{\thefootnote}{\fnsymbol{footnote}}
\begin{document}
\draft
\preprint{DPNU 93-05}
\begin{title}
 Searching for $ T,CP,CPT $
 and $ \DSDQ $ Rule Violations \\
 in the Neutral $ K $ Meson System - a Guide
\end{title}
\author{ M. Hayakawa \footnotemark[2]
     and A.I. Sanda \footnotemark[3] }
\begin{instit}
 Department of Physics, Nagoya University,\\
 Nagoya 464-01, Japan
\end{instit}
\begin{center}
 February 1993
\end{center}
\begin{abstract}
 It is known that the existing experimental limit for
$ CPT $ violation is rather poor; the 10 \% level at best.
 The $ \DSDQ $ rule is tested only up to 2 \%.
 In this note, we discuss the possibility of measuring
$ T, CP, CPT $ and $ \DSDQ $ rule violations in the neutral $ K $
meson system.
 It is shown that not all the symmetry breaking parameters
can be determined at a symmetric $ \phi $ factory.
 Experiments with $ K^0 $ and $ \bar{K}^0 $ beams are needed
to test all the symmetries.
\end{abstract}
\renewcommand{\thefootnote}{\fnsymbol{footnote}}
\footnotetext[2]{e-mail address:e43521a.nucc.cc.nagoya-u.ac.jp}
\footnotetext[3]{e-mail address:sanda.phys.nagoya-u.ac.jp}
\pacs{ }
%
%
%%%%%%%%%%%%%%%%%%%%%%%%%%%%%%%%%%%%%%%%%%%%%%%%%%%%%%%%%%%%%%%%
%                                                              %
%                           Section 1                          %
%                                                              %
%%%%%%%%%%%%%%%%%%%%%%%%%%%%%%%%%%%%%%%%%%%%%%%%%%%%%%%%%%%%%%%%
%
%
\section{ Introduction }
 A clear understanding of $ CP $ violating mechanism is
one of the few things missing in the standard model of
electroweak interactions.
 It is hoped that progress along this direction will lead us
to physics beyond the standard model.
 So far, the only observation of $ CP $ violation is in the \KSM.
 As we expect LEAR at CERN and the $ \phi $ factory at Frascati
\cite{Frascati} to provide further information
on the K meson system,
we feel that a systematic analysis to extract all available
information is timely.
\par
 Once we admit that $ CP $ violation may be a clue to physics beyond
the standard model,
it is important to set up a program which enables us
to study all possible symmetries,
in particular $ CP $, $ CPT $, $ T $ and $ \DSDQ $.
 Within the context of the standard model, only the $ \DSDQ $ current
is allowed.
 The $ \DSMDQ $ current induced by higher order electroweak
corrections is very small \cite{Dib}.
 Similarly, $ CPT $ violating effects are absent
in the standard model of leptons and quarks.
 We do not expect this to change when the standard model is cast
in terms of an effective theory of mesons and baryons.
\par
 We are not content with the existing experimental evidence
for the validity of $ CPT $ and $ \DSDQ $ rule.
 In Sec.\ref{ sec:notation } we present
the experimental situation.
 Experimental limits to the $ \DSDQ $ violation
in the neutral $ K $ meson system
are obtained by measuring the amplitude ratio
\begin{equation}
 x_l = \frac{ A( \bar{K}^0 \rightarrow \pi^- l^+ \nu_l ) }
      { A( K^0 \rightarrow \pi^- l^+ \nu_l ) },
\end{equation}
 for $ l = e $ or $ \mu $.
 Current average values \cite{PDG} are $ \Rp x = 0.006 \pm 0.018 $
and $ \Ip x = -0.003 \pm 0.026 $.
 We can also test $ \DSDQ $ rule violation in the charged $ K $
meson system; the following ratio of widths \cite{PDG}
\begin{equation}
 \frac{ \Gamma( K^+ \rightarrow \pi^+ \pi^+ e^- \bar{\nu}_e ) }
      { \Gamma( K^+ \rightarrow \pi^+ \pi^- e^+ \nu_e ) }
 < 3 \times 10^{-4}
\end{equation}
 yields the upper bound to the $ \DSDQ $ violating process
which is the same order as in the neutral $ K $ meson system,
 $ {\rm i.e.} $ 2 \%.
\par
 Existing experimental evidence for $ CPT $ conservation
is rather poor.
 We see in Sec. \ref{ sec:notation }
that the limit for the strength of $ CPT $ violating interaction
is about 10 \% of $ CP $ violating interaction.
 Also, on the theoretical side, the proof of the $ CPT $ theorem
\cite{Luders}
makes a heavy use of the properties of asymptotic states.
 While it is difficult to construct reasonable $ CPT $ violating
theories, this proof is obsolete, as quarks and gluons in $ QCD $
are not asymptotic states \cite{Kobayashi}.
 This proof should be reexamined.
\par
 We expect that future experiments will improve the accuracy
of both measurements of $ CPT $ and $ \DSDQ $ rule violations.
 So, at this exploratory stage, no assumption about the validity
of these symmetries should be made.
 In this paper we outline a systematic procedure to study
$ CP, T, CPT $ symmetry and $ \DSDQ $ rule.
\par
 We first examine in Sec.\ref{ sec:phi }
the possibility of measuring the violations
of these symmetries at a $ \phi $ factory.
 To our knowledge, a systematic study of
$ CP,\  T, \  CPT,\  \DSDQ $ rule at a $ \phi $ factory is new.
 Similar attempts have been made by Tanner and Dalitz \cite{Tanner}
for LEAR experiments;
 Some of our results also appear in Dunietz, Hauser and Rosner
\cite{Dunietz} who studied the possibility of measuring
$ \epsilon^\prime / \epsilon $ at a $ \phi $ factory;
 Buchanan et.al. \cite{Buchanan} discussed the possibility of
testing $ CP $ and $ CPT $ symmetry at a $ \phi $ factory assuming
$ \DSDQ $.
\par
 We show that separate measurements of $ \Delta, x_{l} $
and $ \bar{x}_l $,
 where these are $ CPT $ and $ \DSDQ $ violating parameters defined
below, cannot be made at a $ \phi $ factory.
 Therefore, experiments with $ K^0 $ and $ \bar{K}^0 $ beam,
such as LEAR experiments,
are necessary for further determination of the parameters.
 Thus in Sec.\ref{ sec:lear } we go on to discuss experiments
with tagged kaon beam.
 Sec.\ref{ sec:discon } is devoted to discussion and conclusion.
\newpage
%
%
%
%
%%%%%%%%%%%%%%%%%%%%%%%%%%%%%%%%%%%%%%%%%%%%%%%%%%%%%%%%%%%%%%%
%                                                             %
%                          Section 2                          %
%                                                             %
%%%%%%%%%%%%%%%%%%%%%%%%%%%%%%%%%%%%%%%%%%%%%%%%%%%%%%%%%%%%%%%
%
%
\section{ Notations and previous experimental results }
\label{ sec:notation }
 In this section, we summarize the notations used throughout
this paper,
and make a few remarks concerning the existing experiments
which test the $ CPT $ and $ \DSDQ $ rule violations.
\par
 We consider the neutral $ K $ meson system
which consists of two states $ \KO $ and $ \KB $.
 These are eigenstates of strangeness $ S $; $ S = 1 $, and $ -1 $
for $ \KO $, and $ \KB $, respectively.
 The time evolution of the arbitrary state
$ \pst = c_1(t) \KO + c_2(t) \KB $ in this system is described
by the equation \cite{GellMann}
\begin{equation}
 \frac{d}{dt}
 \left(
   \begin{array}{c}
     c_1(t) \\
     c_2(t)
   \end{array}
 \right)
 = -
 \left( \frac{\Gamma}{2} + i M \right)
 \left(
   \begin{array}{c}
    c_1(t) \\
    c_2(t)
   \end{array}
 \right),
\label{ eqn:evolution }
\end{equation}
where $ \Gamma $ and $ M $ are $ 2\times 2 $ matrices:
\begin{equation}
\begin{array}{rcl}
 M & = & \left(
 \begin{array}{cc}
  M_{11}   & \,\, M_{12} \\
  M_{12}^* & \,\, M_{22}
 \end{array}
 \right), \\
 \Gamma & = & \left(
 \begin{array}{cc}
  \Gamma_{11}   & \quad \Gamma_{12} \\
  \Gamma_{12}^* & \quad \Gamma_{22}
 \end{array}
 \right).
\end{array}
\end{equation}
 In Appendix A we give the explicit expressions for
the elements $ M_{ij} $ and $ \Gamma_{ij} $
which are valid up to $ H_W^2 $,
where $ H_W $ includes the electroweak interaction,
as well as relations among matrix elements implied
by various discrete symmetries.
 We can decompose the matrix $ \frac{1}{2}\Gamma + i M $ as
\cite{LeeWu}
\begin{equation}
 \frac{\Gamma}{2} + i M
= D {\bf 1} + i( E_1 \sigma_1 + E_2 \sigma_2 + E_3 \sigma_3 ),
\end{equation}
 where $ {\bf 1} $ is the unit matrix and $ \sigma_a ( a=1,2,3 ) $
are Pauli matrices;
\begin{equation}
 \sigma_1 =
 \left(
  \begin{array}{cc}
   \ 0 \ & \ 1 \ \\
   \ 1 \ & \ 0 \
  \end{array}
 \right),
\quad
 \sigma_2 =
 \left(
  \begin{array}{cc}
   \ 0 \ & \ -i \  \\
   \ i \ & \ 0 \
  \end{array}
 \right),
\quad
 \sigma_3 =
 \left(
  \begin{array}{cc}
   \ 1 \ & \ 0 \ \\
   \ 0 \ & \ -1 \
  \end{array}
 \right).
\end{equation}
 The coefficients $ D $ and $ E_a ( a=1,2,3 ) $ are then given by
\begin{equation}
  \begin{array}{l}
    D =
    \displaystyle{
      \frac{1}{4} \left( \Gamma_{11} + \Gamma_{22} \right)
      + \frac{i}{2} \left( M_{11} + M_{22} \right)
    } \\
    E_1 =
    \displaystyle{
      \Rp M_{12} - \frac{i}{2} \Rp \Gamma_{12}
    } \\
    E_2 =
    \displaystyle{
     - \Ip M_{12} + \frac{i}{2} \Ip \Gamma_{12}
    } \\
    E_3 =
    \displaystyle{
     \frac{1}{2} \left( M_{11} - M_{22} \right)
      - \frac{i}{4} \left( \Gamma_{11} - \Gamma_{22} \right)
    }.
  \end{array}
\label{ eqn:DE }
\end{equation}
 We express $ E_a(a=1,2,3) $ in terms of complex polar parameters
$ \theta, \phi $ and $ E $ \cite{LeeWu}:
\begin{equation}
 E_1 = E \sin \theta \cos \phi,\
 E_2 = E \sin \theta \sin \phi,\
 E_3 = E \cos \theta.
\label{ eqn:angle }
\end{equation}
 $ E $ is given by $ E = \sqrt{ E_1^2 + E_2^2 + E_3^2 } $.
\subsection{ Mass eigenstates }
 The eigenvalues of the matrix $ \frac{1}{2} \Gamma + i M $ are
$ \lambda_S = D+iE $ and $ \lambda_L = D-iE $.
 If we denote the mass eigenstates $ \KS(\KL) $
belonging to $ \lambda_S(\lambda_L) $ as:
\begin{equation}
\begin{array}{rcl}
 \KS & = &
\displaystyle{
   \frac{1}{ \sqrt{ \left| p_1 \right|^2 + \left| q_1 \right|^2 } }
             }
\left( p_1 \KO + q_1 \KB \right), \\
 \KL & = &
\displaystyle{
    \frac{1}{ \sqrt{ \left| p_2 \right|^2 + \left| q_2 \right|^2 } }
             }
\left( p_2 \KO - q_2 \KB \right), \\
\label{ eqn:msrl }
\end{array}
\end{equation}
 then we have
\begin{equation}
 \frac{ q_1 }{ p_1 }
= \tan \frac{ \theta }{2} e^{ i \phi }, \quad \quad
 \frac{ q_2 }{ p_2 }
= \cot \frac{ \theta }{2} e^{ i \phi }.
\end{equation}
\subsection{ 2 $ \pi $ decays }
 The amplitudes associated with the $ 2 \pi $ decay modes
are needed in the succeeding section.
 The parameters directly measured in the experiments are
\begin{equation}
\begin{array}{l}
 \eta_{00} =
\displaystyle{
 \frac{ \left< \pi^0 \pi^0 |T| K_L \right> }
 { \left< \pi^0 \pi^0 |T| K_S \right> }
}
= \left| \eta_{00} \right| e^{ i \phi_{00} },
\\
 \eta_{+-} =
\displaystyle{
 \frac{ \left< \pi^+ \pi^- |T| K_L \right> }
 { \left< \pi^+ \pi^- |T| K_S \right> }
}
= \left| \eta_{+-} \right| e^{ i \phi_{+-} }
{}.
\end{array}
\label{ eqn:eta }
\end{equation}
 Using the isospin eigenstates $ \left| ( 2 \pi )_I \right> $
$ ( I = 0, 2 ) $
\begin{equation}
 \begin{array}{l}
  \left< (2\pi)_I^{out} |T| K^0 \right> = A_I e^{ i \delta_I }, \\
  \left< (2\pi)_I^{out} |T| \bar{K}^0 \right> = \bar{A}_I
                          e^{ i \delta_I } \quad ( I=0,2 ),
 \end{array}
\label{ eqn:A }
\end{equation}
 where $ e^{ 2 i \delta_I } \equiv
\left< ( 2\pi )_I^{out} | (2\pi)_I^{in} \right> $
defines the phase shift associated with the strong interaction.
 If one decomposes $ \eta_{00} $ and $ \eta_{+-} $ as
\begin{equation}
\begin{array}{rcl}
 \eta_{+-} & = & \epsilon + \epsilon^\prime,  \\
 \eta_{00} & = & \epsilon - 2 \epsilon^\prime,
\label{ eqn:etaeps }
\end{array}
\end{equation}
 we show in Appendix B that
\begin{equation}
\begin{array}{l}
 \displaystyle{
  \epsilon = \frac{1}{2} \left( 1 - \frac{q_2}{p_2}
             \frac{\bar{A}_0}{A_0} \right)
 }, \\
 \displaystyle{
  \epsilon^\prime
  = \frac{1}{2\sqrt{2}} e^{i(\delta_2 - \delta_0)}
       \left[
        - \omega \left( 1 - \frac{q_2}{p_2} \frac{\bar{A}_0}{A_0}
                 \right)
        + \left( \frac{p_2 A_2 - q_2 \bar{A}_2}{p_2 A_0} \right)
       \right],
 }
\end{array}
\label{ eqn:epsilonepsilonprime }
\end{equation}
 where
\begin{equation}
 \omega \equiv \frac{ \Rp A_2 }{ \Rp A_0 }.
\end{equation}
%
%
%%%%%%%%%%%%%%%%%%%%%%%%%%%%%%%%%%%%%%%%%%%%%%%%%%%%%%%%
%                                                      %
%                     Phase convention                 %
%                                                      %
%%%%%%%%%%%%%%%%%%%%%%%%%%%%%%%%%%%%%%%%%%%%%%%%%%%%%%%%
%
%
\subsection{ Phase convention }
\label{ subsec:Phase }
 As stated before, $ \KO $ and $ \KB $ are eigenstates of
strangeness.
 Since the strangeness quantum number is conserved by strong
interaction,
their relative phase can never be measured.
 So, all observables are independent of the phase transformation
\begin{equation}
\begin{array}{l}
 \KO \rightarrow \KO^\prime = e^{ i\alpha} \KO, \\
 \KB \rightarrow \KB^\prime = e^{-i\alpha} \KB.
\label{ eqn:phasetrans }
\end{array}
\end{equation}
 In general, $ C $ and $ T $ transformations are defined only
up to a phase:
\begin{equation}
\begin{array}{l}
 C \KO = e^{i \phi_C} \KB, \\
 T \KO = e^{i \phi_T} \KO, \quad T \KB = e^{i \phi_T} \KB.
\end{array}
\label{ eqn:phaseofCandT }
\end{equation}
 When the phase tranformation $ \left( \ref{ eqn:phasetrans }
\right) $ is made,
we can adjust $ \phi_C $ and $ \phi_T $ so that, for example,
\begin{equation}
\begin{array}{c}
 C \KO^\prime = - \KB^\prime \\
 T \KO^\prime = \KO^\prime, \quad T \KB^\prime = \KB^\prime.
\end{array}
\end{equation}
 Obviously, the parity operation $ P $ are invariant
under the phase transformation.
 Using the antilinear property of time reversal $ T $,
one can easily show that $ CPT $ operation also remains unchanged.
\par
 We now define a phase convention to fix the phase ambiguity
given in Eq.$ \left( \ref{ eqn:phasetrans } \right) $.
 A widely used phase convention due to Wu and Yang \cite{WuYan} is
\begin{equation}
 \frac{\bar{A}_0}{A_0} = 1.
\label{ eqn:WuYangConv }
\end{equation}
 For this case,
$ q_2 /p_2 = 1 $ and $ E_2 = 0 $ when $ CP $ invariance holds,
since $ M_{12} $ and $ \Gamma_{12} $ are then both real.
 In the standard model, the penguin diagram, which gives
contributions only to
$ A_{0} $ and $ \bar{A}_0 $, contains a phase.
 Thus it naturally leads to
\begin{equation}
 \frac{\bar{A}_2}{A_2} = 1\ {\rm and}\ \frac{\bar{A}_0}{A_0}
= e^{i\lambda}
\end{equation}
 where $ \lambda $ is a small phase.
 Often, it is convenient to choose this phase convention instead of
$ \left( \ref{ eqn:WuYangConv } \right) $.
\par
 We call the Wu-Yang phase convention and other phase conventions
which lead to
\begin{equation}
 \frac{\bar{A}_0}{A_0} \approx 1
\end{equation}
 "physical" phase conventions.
\par
 We give an example of an {\it unphysical} phase convention.
 For an illustrative purpose, consider a world in which both $ CP $
and $ CPT $ are conserved.
 Then the phase invariant quantity
\begin{equation}
 \epsilon
  = \frac{1}{2} \left( 1 - \frac{q_2}{p_2} \frac{\bar{A}_0}{A_0}
                \right)
  = 0.
\end{equation}
\par
 We can choose a wild phase convention so that
\begin{equation}
 \frac{\bar{A}_0}{A_0} = -i.
\end{equation}
 This corresponds to making the transformation
\begin{equation}
\begin{array}{l}
 \KO^\prime = e^{-i \frac{\pi}{4}} \KO, \\
 \KB^\prime = e^{ i \frac{\pi}{4}} \KB,
\end{array}
\end{equation}
where $ \KO^\prime $ and $ \KB^\prime $ follow
the Wu-Yang convention.
 With this phase convention $ q_2 /p_2 = i $ and $ E_1 = 0 $
since both $ M_{12} $ and $ \Gamma_{12} $ are purely imaginary.
 Now the $ CP $ operation $ CP \KO^\prime = \KB^\prime $ corresponds
to
\begin{equation}
 CP \KO = i \KB
\end{equation}
and $ CP $ eigenstates are
\begin{equation}
\begin{array}{l}
 \displaystyle{
  \KS = \frac{1}{1+i} \left( \KO + i \KB \right)
 }, \\
 \displaystyle{
  \KL = \frac{1}{1-i} \left( \KO - i \KB \right)
 }.
\end{array}
\end{equation}
 This implies that if we write $ q_1/p_1 $ and $ q_2/p_2 $ as
\begin{equation}
 \frac{1-\epsilon_1}{1+\epsilon_1} \equiv \frac{q_1}{p_1},\ \
 \frac{1-\epsilon_2}{1+\epsilon_2} \equiv \frac{q_2}{p_2},
\end{equation}
then
$ \epsilon_1 = 1/\epsilon_2 = \frac{1-i}{1+i} $.
 We can, therefore, see that
neither nonvanishing $ \Ip{M_{12}},\Ip{\Gamma_{12}} $ nor
$ \epsilon_{1,2} $
automatically lead to $ CP $ violation.
\par
 We know experimentally that $ CP $ violating effects are very small
in the $ K $ meson system.
 Thus it makes sense to confine ourselves to a class of
phase convention such that small $ \epsilon $
and $ \epsilon^\prime $ imply small $ \Ip{M_{12}} $
and $ \Ip{\Gamma_{12}} $.
 In addition, we take
\begin{eqnarray}
 && C \KO = - \KB \\
 && \left\{
     \begin{array}{l}
      T \KO = \KO \\
      T \KB = \KB
     \end{array}
    \right.,\
 \left\{
     \begin{array}{l}
      P \KO = - \KO \\
      P \KB = - \KB.
     \end{array}
    \right. \\
 && CPT \KO = \KB,\ CP \KO = \KB
\end{eqnarray}
 From Appendix A and Eq.$ \left( \ref{ eqn:DE } \right) $
we can see that for this class of physical phase conventions
\begin{itemize}
 \item $ CPT $ conservation $ \Longrightarrow $ $ \cos \theta = 0 $
 \item $ CP $ conservation $ \Longrightarrow $ $ \cos \theta = 0,
     \  \sin \phi = 0 $
 \item $ T $ conservation $ \Longrightarrow $ $ \sin \phi = 0. $
\end{itemize}
 Note that each arrow points only to one direction.
\par
 We list below the approximate expressions for various quantities
which are valid when we restrict ourselves
to the physical phase conventions.
 First note that
\begin{equation}
 E_1^2 + E_2^2 = \left| M_{12} \right|^2
+ \frac{1}{4} \left| \Gamma_{12} \right|^2
+ i \Rp\left( M_{12} \Gamma_{12}^* \right)
\end{equation}
 is independent of phase convention.
 For the physical phase conventions,
 we expect the magnitude of $ E_2 $ and $ E_3 $ to be much smaller
than $ E_1 $ since the former is at least proportional to
one of the $ CP,T $, or $ CPT $ violating parameters.
 So we get $ E \simeq E_1 $.
 Picking up their real and imaginary parts, we have
\begin{equation}
 \begin{array}{l}
  m_S =
  \displaystyle{
    \frac{1}{2} \left( M_{11} + M_{22} + 2 \Rp M_{12} \right)
  }, \\
  m_L =
  \displaystyle{
    \frac{1}{2} \left( M_{11} + M_{22} - 2 \Rp M_{12} \right)
  }, \\
  \Gamma_S =
  \displaystyle{
    \frac{1}{2} \left( \Gamma_{11} + \Gamma_{22} + 2 \Rp \Gamma_{12}
                \right)
  }, \\
 \Gamma_L =
  \displaystyle{
    \frac{1}{2} \left( \Gamma_{11} + \Gamma_{22} - 2 \Rp \Gamma_{12}
                \right)
  },
\end{array}
\label{ eq:massvalue }
\end{equation}
 where $ \ws (\wl) $, and $ m_S (m_L) $ are the total decay width,
and the mass of $ K_S (K_L) $.
 The parameter $ \epsilon_1 ( \epsilon_2 ) $ represents the small
deviation of the mass eigenstate $ K_S ( K_L ) $
from the $ CP $ eigenstate $ K_+ ( K_- ) $ :
\begin{equation}
\begin{array}{rcl}
 \left| K_+ \right>
   & = & \displaystyle{ \frac{ \KO + \KB }{ \sqrt{2} } }, \\
 \left| K_- \right>
   & = & \displaystyle{ \frac{ \KO - \KB }{ \sqrt{2} } }.
\end{array}
\end{equation}
 When $ CPT $ invariance is a good symmetry, one finds
\begin{equation}
 \frac{ q_1 }{ p_1 }
= \frac{ q_2 }{ p_2 }
= e^{ i \phi },
\end{equation}
 so that
\begin{equation}
 \epsilon_1 = \epsilon_2 = -i \tan \frac{ \phi }{2}.
\end{equation}
\par
 It is customary to write $ \epsilon_1 $ and $ \epsilon_2 $ as
\cite{Tanner,Buchanan}
\begin{equation}
\begin{array}{rcl}
 \epsilon_1 & = & \epsilon_0 + \Delta, \nonumber \\
 \epsilon_2 & = & \epsilon_0 - \Delta,
\label{ eqn:epsDelta }
\end{array}
\end{equation}
 where, if $ \epsilon_{0} $ and $ \Delta $ are small,
\begin{equation}
\begin{array}{rcl}
 \epsilon_0 & \equiv & -i \displaystyle{ \frac{ \phi }{2} }, \\
 \Delta   & \equiv & - \displaystyle{ \frac{1}{2}
\left( \theta - \frac{ \pi }{2} \right) }.
\label{ eqn:cor }
\end{array}
\end{equation}
 Above we have taken $ \theta = \frac{\pi}{2} $
when $ \cos\theta = 0 $,and $ \phi = 0 $ when $ \sin\phi = 0 $.
 It is instructive to express these parameters
in terms of $ M_{ij} $ and $ \Gamma_{ij} $.
 Using Eqs.
$ \left( \ref{ eqn:DE } \right), \left( \ref{ eqn:angle } \right) $
and $ \left( \ref{ eqn:cor } \right) $, we get \cite{Tanner}
\begin{equation}
 \begin{array}{l}
  \displaystyle{
   \epsilon_0 =
    \frac{ - \Ip M_{12} + \frac{i}{2} \Ip \Gamma_{12} }
         { \frac{1}{2} \Delta\Gamma - i \Delta m }
  }, \\
  \displaystyle{
   \Delta = \frac{1}{2}
             \frac{ i( M_{11} - M_{22} )
             + \frac{1}{2} ( \Gamma_{11} - \Gamma_{22} ) }
               { \frac{1}{2} \Delta \Gamma - i \Delta m }
  },
 \end{array}
\label{ eqn:epsDeltaMass }
\end{equation}
 where $ \Delta m = m_L - m_S $,
and $ \Delta \Gamma \equiv \ws - \wl $.
 From Eq. $ \left( \ref{ eqn:cor } \right) $,
or using Eq. $ \left( \ref{ eqn:epsDeltaMass } \right) $
and Appendix A, we can see that
\begin{itemize}
 \item $ \epsilon_0 \neq 0 $ $ \Longrightarrow $ $ CP $
and $ T $ violating, but may be $ CPT $ conserving,
 \item $ \Delta \neq 0 $ $ \Longrightarrow $ $ CP $
and $ CPT $ violating, but may be $ T $ conserving.
\end{itemize}
 Each discrete symmetry reduces the freedom of $ A_I $
and $ \bar{A}_I $ as follows;
\begin{itemize}
 \item $ CPT $ conservation $ \Longrightarrow $ $ A_I = \bar{A}_I^* $
 \item $ CP $ conservation  $ \Longrightarrow $ $ A_I = \bar{A}_I $
 \item $ T $ conservation   $ \Longrightarrow $
  $ \Ip A_I = 0 = \Ip \bar{A}_I $.
\end{itemize}
 If we regard $ \Ip{A_I} /\Rp{A_I}, \Ip{\bar{A}_I} /\Rp{A_I} $
and $ ( 1 - \Rp{\bar{A}_I} /\Rp{A_I} ) $ as so small
that they are the same order quantities
as $ \epsilon_1, \epsilon_2 $, we have
\begin{eqnarray}
 \epsilon &=& \epsilon_2
+
\displaystyle{
 \frac{1}{2}
 \left( 1 -
      \frac{ \Rp \bar{A}_0 }{ \Rp A_0 }
 \right)
 + \frac{i}{2}
 \left(
      \frac{ \Ip A_0 }{ \Rp A_0 }
      - \frac{ \Ip \bar{A}_0 }{ \Rp A_0 }
 \right)
}, \nonumber \\
 \epsilon^\prime &=&
\displaystyle{
 \frac{1}{2 \sqrt{2}} \omega e^{ i( \delta_2 - \delta_0 ) }
 \biggl[ \left\{ \left( 1 - \frac{ \Rp \bar{A}_2 }{ \Rp A_2 } \right)
               - \left( 1 - \frac{ \Rp \bar{A}_0 }{ \Rp A_0 } \right)
        \right\}
} \nonumber \\
 & & \quad \quad \quad \quad \quad +
\displaystyle{
 i \left\{ \left( \frac{ \Ip A_2 }{ \Rp A_2 }
                          - \frac{ \Ip \bar{A}_2 }{ \Rp A_2 }
           \right)
                      - \left( \frac{ \Ip A_0 }{ \Rp A_0 }
                           - \frac{ \Ip \bar{A}_0 }{ \Rp A_0 }
                        \right)
   \right\} \biggl]
},
\label{ eqn:epsilonprime }
\end{eqnarray}
\par
 Following to the Wu-Yang phase convention $ \Ip{A_0} = 0 $
\cite{WuYan}
and using Eq.$ \left( \ref{ eqn:epsDeltaMass } \right) $
we can rewrite $ \epsilon $
in Eq.$ \left( \ref{ eqn:epsilonprime } \right) $ as follows;
\begin{eqnarray}
 \epsilon &=&
  \biggl[
  \frac{ -\Ip{M_{12}} + \frac{1}{2} \Ip{\Gamma_{12}} }
       { \frac{1}{2} \Delta\Gamma - i\dm }
 - \frac{i}{2} \frac{\Ip\bar{A}_0}{A_0}
 \biggl]
- \biggl[
 \frac{1}{2}
    \frac{ i( M_{11}-M_{22} )
      + \frac{1}{2} ( \Gamma_{11} - \Gamma_{22} ) }
         { \frac{1}{2} \Delta\Gamma - i \dm }
 \nonumber \\
 & &
+ \frac{1}{2} \left( 1 - \frac{\Rp{\bar{A}_0}}{A_0} \right) \biggl].
\end{eqnarray}
 In this equation each term in the second bracket is
$ CPT $ violating.
 Define the so-called superweak phase $ \phi_{SW} $ as
\begin{equation}
 e^{ -i \phi_{SW} } \equiv
  \frac{ \frac{1}{2} \Delta\Gamma - i\dm }
 { \sqrt{ (\dm)^2 + \left( \frac{1}{2} \Delta\Gamma \right)^2 } }.
\end{equation}
 Then we get
\begin{eqnarray}
 \epsilon \cdot e^{ -i\phi_{SW} } & = &
 \displaystyle{
   \frac{ -\Ip{M_{12}} + \frac{i}{2} \Ip{\Gamma_{12}} }
      { \sqrt{ (\dm)^2 + \left( \frac{1}{2}\Delta\Gamma \right)^2 } }
   - \frac{1}{2}
     \frac{ i( M_{11}-M_{22} )
            + \frac{1}{2} (\Gamma_{11} - \Gamma_{22}) }
     { \sqrt{ (\dm)^2 + \left( \frac{1}{2} \Delta\Gamma \right)^2 } }
 } \nonumber \\
 & & + \displaystyle{
        \frac{1}{2} \left( 1 - \frac{\bar{A}_0}{A_0} \right)
        \frac{ \frac{1}{2} \Delta\Gamma - i\dm }
             { \sqrt{ (\dm)^2 + \left( \frac{1}{2} \Delta\Gamma
               \right)^2 } }
       }.
\nonumber
\end{eqnarray}
 By taking its imaginary part we have
\begin{eqnarray}
 \Ip( \epsilon \cdot e^{ -i\phi_{SW} } ) &=&
  \displaystyle{
   \frac{\Delta\Gamma}{2\dm} \sin\phi_{SW}
   \biggl[
    \frac{\Ip\Gamma_{12}}{\Delta\Gamma}
    - \frac{ M_{11} - M_{22} }{ \Delta\Gamma }
  } \nonumber \\
 & & \quad \quad
  \displaystyle{
   - \frac{\dm}{\Delta\Gamma}
       \left( 1 - \frac{\Rp\bar{A}_0}{A_0} \right)
   - \frac{1}{2} \frac{\Ip{\bar{A}_0}}{A_0}
  \biggl].
  }
\label{ eqn:expconst }
\end{eqnarray}
\par
 Now we discuss the exprerimental constraint to $ CPT $ violation.
 Experiment \cite{Carosi} measures $ \eta_{+-} $ and $ \eta_{00} $.
 Then the value of the quantity
\begin{eqnarray}
 \Ip( \epsilon \cdot e^{ -i \phi_{SW} } )
& \simeq &
\displaystyle{
 \left| \eta_{+-} \right|
 \left( \frac{2}{3} \phi_{+-} + \frac{1}{3} \phi_{00} - \phi_{SW}
\right)
} \nonumber \\
 & = &
(1.3 \pm 0.8) \times 10^{-4},
\end{eqnarray}
 which is obtained using the fact that
$ \left| \eta_{+-} \right| \simeq \left| \eta_{00} \right| \simeq
\left| \epsilon \right| $, and $ \phi_{+-} \simeq \phi_{00} $,
can be determined.
 If we define $ \xi $ as
\begin{equation}
 \xi \equiv \arg( \Gamma_{12} A_0 \bar{A}_0^* )
  \simeq 2 \frac{ \Ip\Gamma_{12} }{ \Delta\Gamma }
          - \frac{ \Ip{\bar{A}_0} }{A_0},
\end{equation}
 we can deduce from Eq.$ \left( \ref{ eqn:expconst } \right) $
\begin{equation}
 \frac{\xi}{2} - \frac{M_{11}- M_{22}}{\Delta\Gamma}
- \frac{\dm}{\Delta\Gamma} \left( 1 - \frac{\Rp\bar{A}_0}{A_0}
                           \right)
= \left( 1.8 \pm 1.1 \right) \times 10^{-4},
\label{ eqn:CPTconst }
\end{equation}
 which is the expression obtained by Lavoura \cite{Lavoura}.
 Note that $ \xi \not= 0 $ even when $ CPT $ invariance holds.
 As pointed out by Lavoura \cite{Lavoura}
and earlier by Lee and Wu \cite{LeeWu}, the above experimental result
does not supply $ CPT $ test as long as the value of $ \xi $
is not determined with some accuracy.
 The contributions to $ \xi $
from semileptonic decay modes($ \xi( \pi l \nu ) $),
and from $ 3 \pi $ modes($ \xi( 3 \pi ) $) can be as large as
\cite{LeeWu,Lavoura}
\begin{equation}
\begin{array}{c}
 \xi(\pi l \nu) = 1.9 \times 10^{-4}, \\
 \xi(3 \pi) = 3.8 \times 10^{-4}.
\end{array}
\end{equation}
 Hence from Eq.$ \left( \ref{ eqn:CPTconst } \right) $
these imply that
\begin{equation}
 39.5^{\circ} < \arg\epsilon < 47.4^{\circ}.
\end{equation}
 Comparing $ \Ip\left( \epsilon \cdot e^{-i\phi_{SW}} \right) $
with the size of $ \epsilon $, and taking its theoretical error
into account,
we can say optimistically that $ CPT $ is good up to 10 \%.
\subsection{ Semileptonic decays }
 The amplitudes for semileptonic decays are given as follows
\cite{Tanner};
\begin{equation}
\begin{array}{rcl}
 \left< \pi^- l^+ \nu_l |T| K^0 \right> & = & F_l ( 1 - y_l ), \\
 \left< \pi^- l^+ \nu_l |T| \bar{K}^0 \right>
         & = & x_l F_l ( 1 - y_l ), \\
 \left< \pi^+ l^- \bar{\nu}_l |T| K^0 \right> & = &
\bar{x}^*_l F^*_l ( 1 + y^*_l ), \\
 \left< \pi^+ l^- \bar{\nu}_l |T| \bar{K}^0 \right> & = &
            F^*_l ( 1 + y^*_l )
\quad \quad ( l = e,\, \mu ).
\end{array}
\end{equation}
 These match with the physical phase convention
since the meaning of each parameter is characterized by
\begin{itemize}
 \item $ \DSDQ $ rule $ \Longrightarrow $ $ x_l = \bar{x}_l = 0 $
 \item $ CPT $ conservation $ \Longrightarrow $ $ y_l = 0,
        x_l = \bar{x}_l $
 \item $ CP $ conservation $ \Longrightarrow $ $ x_l = \bar{x}_l^*,
      F_l = F_l^*, y_l = - y_l^* $.
\end{itemize}
\subsection{ Probability densities for semileptonic decays
for $ K^0 $ and $ \bar{K}^0 $ beams }
 Before making contact with experiments previously performed
for the $ x $ and $ \bar{x}_l $,
we list here several time dependent probability functions.
\par
 Let $ \KT $ be a state which starts out as $ \KO $ at $ t = 0 $,
and $ \KBT $ be a state which starts out as $ \KB $ at $ t = 0 $.
 Then the time dependence of $ \KT $ and $ \KBT $ are given as:
\begin{equation}
\begin{array}{c}
 \displaystyle{
  \KT = \frac{1}{ p_1 q_2 + p_2 q_1 }
   \Bigl[ \left( p_1 q_2 \KO + q_1 q_2 \KB \right)
                          e^{ -i m_S t - \frac{\ws}{2} t }
 } \\
 \quad \quad
 \displaystyle{
  + \left( p_2 q_1 \KO - q_1 q_2 \KB \right)
                             e^{ -i m_L t - \frac{\wl}{2} t }
  \Bigl]
 }, \\
 \\
 \displaystyle{
  \KBT = \frac{1}{ p_1 q_2 + p_2 q_1 }
   \Bigl[ \left( p_1 p_2 \KO + p_2 q_1 \KB \right)
                               e^{ -i m_S t - \frac{\ws}{2} t }
 } \\
 \quad \quad
 \displaystyle{
  - \left( p_1 p_2 \KO - p_1 q_2 \KB \right)
                               e^{ -i m_L t - \frac{\wl}{2} t }
  \Bigl]
 }.
\end{array}
\end{equation}
\par
 The following quantities are the probabilities at time $ t $
for the states $ \KT $ and $ \KBT $ to decay into $ \pi^- l^+ \nu_l $
or $ \pi^+ l^- \bar{\nu}_l $ \cite{Tanner};
\begin{eqnarray}
 \PA & \equiv &
     \left| \left< \pi^- l^+ \nu_l |T| K(t) \right> \right|^2
\nonumber \\
 & = &
  \displaystyle{
   \frac{ \left| F_l \right|^2 }
        { \left| p_1 q_2 + p_2 q_1 \right|^2 }
         \left| 1 - y_l \right|^2
  } \nonumber \\
 & & \quad \quad \times
       \displaystyle{
         \left| p_1 q_2 \left( 1 + \frac{q_1}{p_1} x_l \right)
                            e^{ -i m_S t - \frac{\ws}{2} t }
              + p_2 q_1 \left( 1 - \frac{q_2}{p_2} x_l \right)
                            e^{ -i m_L t - \frac{\wl}{2} t }
         \right|^2
       } \\
\label{ eqn:painv }
\end{eqnarray}
\begin{eqnarray}
 \BA & \equiv & \left| \left< \pi^- l^+ \nu_l |T| \bar{K}(t) \right>
\right|^2 \nonumber \\
 & = & \displaystyle{
        \left| F_l \right|^2
        \frac{ \left| p_1 p_2 \right|^2 }{ \left| p_1 q_2 + p_2 q_1
               \right|^2 }
        \left| 1 - y_l \right|^2
       } \nonumber \\
 & & \quad \quad \times
     \displaystyle{
      \left|
        \left( 1 + \frac{q_1}{p_1} x_l \right)
          e^{ -i m_S t - \frac{\ws}{2} t }
      - \left( 1 - \frac{q_2}{p_2} x_l \right)
          e^{ -i m_L t - \frac{\wl}{2} t }
      \right|^2
     }
\label{ eqn:bainv }
\end{eqnarray}
\begin{eqnarray}
 \PL & \equiv & \left| \left< \pi^+ l^- \bar{\nu}_l |T| K(t) \right>
\right|^2 \nonumber \\
 & = & \displaystyle{
        \left| F_l \right|^2
        \frac{ \left| q_1 q_2 \right|^2 }{ \left| p_1 q_2 + p_2 q_1
               \right|^2 }
        \left| 1 + y_l^* \right|^2
       } \nonumber \\
 & & \quad \quad \times
     \displaystyle{
      \left|
        \left( 1 + \frac{p_1}{q_1} \bar{x}_l^* \right)
             e^{ -i m_S t - \frac{\ws}{2} t }
      - \left( 1 - \frac{p_2}{q_2} \bar{x}_l^* \right)
             e^{ -i m_L t - \frac{\wl}{2} t }
      \right|^2
     }
\label{ eqn:plinv }
\end{eqnarray}
\begin{eqnarray}
 \BL & \equiv & \left| \left< \pi^+ l^- \bar{\nu}_l |T| \bar{K} (t)
                \right> \right|^2 \nonumber \\
 & = & \displaystyle{
         \frac{ \left| F_l \right|^2 }
              { \left| p_1 q_2 + p_2 q_1 \right|^2 }
         \left| 1 + y_l^* \right|^2
       } \nonumber \\
 & & \quad \quad \times
       \displaystyle{
         \left| p_2 q_1
            \left( 1 + \frac{p_1}{q_1} \bar{x}_l^* \right)
                   e^{ -i m_S t - \frac{\ws}{2} t }
         + p_1 q_2 \left( 1 - \frac{p_2}{q_2} \bar{x}_l^* \right)
                   e^{ -i m_L t - \frac{\wl}{2} t }
         \right|^2
       }
\label{ eqn:blinv }
\end{eqnarray}
 where $ \wt \equiv \ws + \wl $.
 We should note that we can measure at best phase-convention
independent quantities:
\begin{equation}
 \displaystyle{
  1 + \frac{q_1}{p_1} x_l,\quad 1 - \frac{q_2}{p_2} x_l,\quad
  1 + \frac{p_1}{q_1} \bar{x}_l,\quad 1 - \frac{p_2}{q_2} \bar{x}_l,
 }
\end{equation}
and so on.
 However, as already assumed in previous several subsections,
we carry out the computation up to the first order
in $ \epsilon_0, \Delta, x_l, \bar{x}_l $ and $ y_l $
based on the physical phase convention.
 While it is trivial to continue to perform the general computation,
we found it neither useful nor instructive at this stage.
 Then the equations encountered previously become
\begin{equation}
\begin{array}{rcl}
 \KT & = &
\displaystyle{ \frac{1}{2} } \Bigl[
\left\{ ( 1 + 2 \Delta ) \KO + ( 1 - 2 \epsilon_0 ) \KB \right\}
e^{ - i m_S t - \frac{\ws}{2} t }  \\
 & & \quad + \left\{ ( 1 - 2 \Delta ) \KO - ( 1 - 2 \epsilon_0 ) \KB
\right\} e^{ - i m_L t - \frac{\wl}{2} t} \Bigl], \\
 & & \\
 \KBT & = &
\displaystyle{ \frac{1}{2} } \Bigl[
\left\{ ( 1 + 2 \epsilon_0 ) \KO + ( 1 - 2 \Delta ) \KB \right\}
e^{ - i m_S t - \frac{\ws}{2} t }  \\
 & & \quad - \left\{ ( 1 + 2 \epsilon_0 ) \KO - ( 1 + 2 \Delta )
\KB \right\} e^{ - i m_L t - \frac{\wl}{2} t } \Bigl],
\label{ eqn:time }
\end{array}
\end{equation}
and
\begin{eqnarray}
 \PA
& = & \left| \displaystyle{ \frac{ F_l }{2} } \right|^2
\Bigl[ \left( 1 - 2 \Rp y_l - 2 \Rp( 2 \Delta + x_l ) \right)
e^{ -\wl t } \nonumber \\
 & & + \left( 1 - 2 \Rp y_l + 2 \Rp( 2 \Delta + x_l ) \right)
e^{ -\ws t }
\nonumber \\
 & & + 2 e^{ -\frac{\wt}{2} t } \left\{ ( 1 - 2 \Rp y_l )
\cos( \dm t ) - 2 \Ip( 2 \Delta + x_l ) \sin( \dm t )
       \right\} \Bigl],
\label{ eqn:pa }
\end{eqnarray}
\begin{eqnarray}
 \BA
 & = & \left| \displaystyle{ \frac{ F_l }{2} } \right|^2
\Bigl[ \left( 1 + 2 \Rp( 2 \epsilon_0 - y_l ) - 2 \Rp x_l \right)
        e^{ -\wl t }       \nonumber \\
 & & + \left( 1 + 2 \Rp( 2 \epsilon_0 - y_l ) + 2 \Rp x_l \right)
    e^{ -\ws t }           \nonumber \\
 & & - 2 e^{ -\frac{\wt}{2} t }
          \left\{ ( 1 + 2 \Rp( 2 \epsilon_0 - y_l ) ) \cos( \dm t )
- 2 \Ip x_l \sin( \dm t ) \right\} \Bigl],
\label{ eqn:ba }
\end{eqnarray}
\begin{eqnarray}
 \PL
 & = & \left| \displaystyle{ \frac{ F_l }{2} } \right|^2
\Bigl[ \left( 1 - 2 \Rp( 2 \epsilon_0 - y_l ) - 2 \Rp \bar{x}_l)
        \right) e^{ -\wl t }
\nonumber \\
 & & + \left( 1 - 2 \Rp( 2 \epsilon_0 - y_l ) + 2 \Rp \bar{x}_l
       \right) e^{ -\ws t }
\nonumber \\
 & & - 2 e^{ -\frac{\wt}{2} t }
      \left\{ ( 1 - 2 \Rp( 2 \epsilon_0 - y_l ) ) \cos( \dm t )
+ 2 \Ip \bar{x}_l \sin( \dm t ) \right\} \Bigl],
\label{ eqn:pl }
\end{eqnarray}
\begin{eqnarray}
 \BL
 & = & \left| \displaystyle{ \frac{ F_l }{2} } \right|^2
\Bigl[ \left( 1 + 2 \Rp y_l + 2 \Rp( 2 \Delta - \bar{x}_l ) \right)
e^{ -\wl t } \nonumber \\
 & & + \left( 1 + 2 \Rp y_l - 2 \Rp( 2 \Delta - \bar{x}_l ) \right)
e^{ -\ws t } \nonumber \\
 & & + 2 e^{ -\frac{\wt}{2} t }\left\{ ( 1 + 2 \Rp y_l )
\cos( \dm t )
+ 2 \Ip( 2 \Delta + \bar{x} _l ) \sin( \dm t ) \right\} \Bigl],
\label{ eqn:bl }
\end{eqnarray}
\par
 Now, $ CPT $ violation is characterized by nonvanishing
$ x_l - \bar{x}_l $ or $ y_l $, representing the violations
in the leptonic decay amplitudes,
and by $ \Delta $, representing the violation
in the $ K^0 $ - $ \bar{K}^0 $ mass matrix.
 Note that only the combination of $ 2 \Delta + x_l $,
and $ 2 \Delta - \bar{x}_l^* $
can be determined from
Eq.$ \left( \ref{ eqn:pa } \right) $,
and Eq.$ \left( \ref{ eqn:bl } \right) $, respectively.
 Thus the test of $ CPT $ symmetry
( nonvanishing $ \Delta $ or $ x_l - \bar{x}_l $ )
can not be performed
without further information on the $ \DSDQ $ rule.
\subsection{ The experimental situation for $ \DSDQ $ violation }
 First note that according to the Review of Particle Properties
\cite{PDG} there is no new experimental results
on the $ \DSDQ $ rule since 1973.
 All existing experiments observe the quantities
corresponding to $ \PA $ and $ \PL $ for the determination of
$ x $ \cite{DSDQ}.
 Parameters which account for $ CP,T $ and $ CPT $ violations
were ignored.
 In this limit, Eqs. $ \left( \ref{ eqn:pa } \right) $,
and $ \left( \ref{ eqn:pl } \right) $ reduces to \cite{DSDQ}
\begin{eqnarray}
 \PA & = & \left| \displaystyle{ \frac{F_l}{2} } \right|^2 \Bigl[
( 1 - 2 \Rp x_l ) e^{ -\wl t } + ( 1 + 2 \Rp x_l ) e^{ -\ws t }
\nonumber \\
 & & + 2 e^{ -\frac{\wt}{2} t } \cos( \dm t )
- 4 \Ip x_l e^{ -\frac{\wt}{2} t } \sin( \dm t ) \Bigl],
\label{ eqn:cppa }
\end{eqnarray}
\begin{eqnarray}
 \PL & = & \left| \displaystyle{ \frac{F_l}{2} } \right|^2 \Bigl[
( 1 - 2 \Rp x_l ) e^{ -\wl t }
+ ( 1 + 2 \Rp x_l ) e^{ -\ws t }
\nonumber \\
 & & - 2 e^{ -\frac{\wt}{2} t } \cos( \dm t )
- 4 \Ip x_l e^{ -\frac{\wt}{2} t } \sin( \dm t ) \Bigl],
\label{ eqn:cppl }
\end{eqnarray}
respectively.
 The statistical average of all experiments gives \cite{PDG}
\begin{equation}
 \begin{array}{rcl}
  \Rp x & = & 0.006 \pm 0.018, \\
  \Ip x & = & -0.003 \pm 0.026,
 \end{array}
\end{equation}
( $ x^\prime $s for $ K_{e3} $ and $ K_{\mu 3} $ are combined )
while individual experiments have error of order $ 0.03 $.
 At the present experimental accuracy,
 $ CP,T $ and $ CPT $ violation can be neglected
in the determination of $ x $.
 However as suggested in the {\rm Ref.}\cite{LEAR},
the LEAR experiment may reach
$ 6 \times 10^{-4} $ for $ \Rp x $
and $ 7 \times 10^{-4} $ for $ \Ip x $.
 At this level it is important to keep all the parameters.
%
%
%
%
%
%
%%%%%%%%%%%%%%%%%%%%%%%%%%%%%%%%%%%%%%%%%%%%%%%%%%%%%%%%%%%%%%%
%                                                             %
%                           section 3                         %
%                                                             %
%%%%%%%%%%%%%%%%%%%%%%%%%%%%%%%%%%%%%%%%%%%%%%%%%%%%%%%%%%%%%%%
%
%
\section{ Measurements at a $ \phi $ factory }
\label{ sec:phi }
 In this section, we discuss how to extract the values of parameters
chracterizing the $ K^0 $ - $ \bar{K}^0 $ system
at a $ \phi $ factory.
 A $ \phi $ factory is expected to produce a large number
of $ K^0 $ and $ \bar{K}^0 $
through the decay $ \phi \rightarrow K^0 \bar{K}^0 $.
 Because strong interaction conserves $ CP $
and $ \phi $ is $ CP $ odd, $ K^0 \bar{K}^0 $ state assumes the form
\begin{equation}
 \frac{1}{\sqrt{2}}
\left\{
\left| K^0 \right>_{ \bf p } \left| \bar{K}^0  \right>_{ - \bf p }
- \left| \bar{K}^0 \right>_{ \bf p } \left| K^0 \right>_{ - \bf p }
\right\}.
\end{equation}
 Here $ {\bf p} $ is the space momentum of one of
the two $ K $ mesons at the $ \phi $ rest frame.
 Each $ K $ meson evolves in time
according to Eq. $ \left( \ref{ eqn:time } \right) $.
 Let one of them decay to the final state
$ \left| f_1 \right> $ at time $ t_1 $,
 and the other to $ \left| f_2 \right> $ at time $ t_2 $.
 The amplitude for such a process is given by
\begin{eqnarray}
 A( f_1, t_1 ; f_2, t_2 ) & = &
\displaystyle{
 \frac{1}{\sqrt{2}} \Bigl[
\left< f_1 |T| K(t_1) \right> \left< f_2 |T| \bar{K} (t_2) \right>
}
\nonumber \\
 & & \quad \quad
- \left< f_1 |T| \bar{K} (t_1) \right> \left< f_2 |T| K(t_2) \right>
\Bigl].
\nonumber \\
\end{eqnarray}
 The relative time probability distribution function,
which is definedby
\begin{equation}
 \left| A( \dt; f_1, f_2 ) \right| \equiv
\frac{1}{2} \int_{\left| \Delta t \right|}^\infty d( t_1 + t_2 )
\left| A( f_1, t_1; f_2, t_2 ) \right |^2,
\label{ eqn:reltime }
\end{equation}
with $ \Delta t = t_1 - t_2 $,
plays the central role in our foregoing analysis.
 This quantity is especially useful as the determination
of $ t_1 + t_2 $ at a $ \phi $ factory is often accompanied
by a large error.
\par
 Since we are interested in observing the violations of
the $ \DSDQ $ rule,
$ CPT $ and $ CP $ symmetry, we shall deal with
$ f_1, f_2 = \pi^- l^+ \nu_l, \pi^+ l^- \bar{\nu}_l, \pi^+ \pi^-,
\pi^0 \pi^0, $
etc.
 For completeness we give the expressions for
$ \left| A( f_1, t_1; f_2, t_2 ) \right|^2 $,
and for the relative time
probability distribution functions in Appendix \ref{ app:prob },
and in Appendix \ref{ app:relprob }, respectively.
 They are also useful when various efficiency and errors
must be folded in to the experimental data.
\par
 As claimed in the introduction, we focus our attention
to various asymmetries,
each of which are obtained by taking the ratio of the difference
and the sum of two different observables.
we list the time integrated asymmetries in Sec.\ref{ subsec:cas },
and $ \dt $ dependent asymmetries in Sec.\ref{ subsec:tas }.
%
%
%%%%%%%%%%%%%%%%%%%%%%%%%%%%%%%%%%%%%%%%%%%%%%%%%%%%%%%%%%%%%%
%                                                            %
%                         Subsection 3.1                     %
%                                                            %
%%%%%%%%%%%%%%%%%%%%%%%%%%%%%%%%%%%%%%%%%%%%%%%%%%%%%%%%%%%%%%
%
\subsection{ Time integrated asymmetries }
\label{ subsec:cas }
\par
 In the $ f_1 = \pi^+ \pi^-,\ f_2 = \pi^0 \pi^0 $ mode,
\begin{eqnarray}
 \lefteqn{ {\cal A}^{ \pi^+ \pi^-, \pi^0 \pi^0 } }
\nonumber \\
 & \equiv &
\displaystyle{
 \frac{
  \int_0^\infty d(\dt) \left| A( \dt; \pi^+ \pi^-, \pi^0 \pi^0 )
                       \right|^2
 - \int_{-\infty}^0 d(\dt) \left| A( \dt; \pi^+ \pi^-, \pi^0 \pi^0 )
                           \right|^2
 }
 {
  \int_0^\infty d(\dt) \left| A( \dt; \pi^+ \pi^-, \pi^0 \pi^0 )
                       \right|^2
 + \int_{-\infty}^0 d(\dt) \left| A( \dt; \pi^+ \pi^-, \pi^0 \pi^0 )
                           \right|^2
 }
}
\nonumber \\
 & = & 3 \Rp ( \epsilon^\prime /\epsilon ),
\label{ eqn:pipi }
\end{eqnarray}
 which was obtained in Ref. \cite{Dunietz,Buchanan,KEK}
remains obviously unaffected
even when  the possibility of $ \DSDQ $ rule violation is taken
into account,
since they appear only in the semileptonic decay channel.
 In obtaining Eq.$ \left( \ref{ eqn:pipi } \right) $, we have used
$ \wl/\ws \sim 10^{-3} $.
 $ \cal{A}^{\pi^+ \pi^-,\, \pi^0 \pi^0} $ reflects the difference
$ \epsilon^\prime $ between $ \eta_{+-} $ and $ \eta_{00} $
( see Eq. $ \left( \ref{ eqn:etaeps } \right) $ ).
Ref. \cite{KEK} notices that the measurement of the double ratio
$ R \equiv \left| \eta_{+-}/\eta_{00} \right|^2
\simeq 1 + 6 \Rp( \epsilon^\prime /\epsilon ) $
enables us to determine $ \Rp( \epsilon^\prime /\epsilon ) $
more accurately than  any other observable such as
$ {\cal A}^{\pi^+ \pi^- \, \pi^0 \pi^0} $,
since the cancellation of most of systematic errors can be reached.
\par
 For $ f_1 = \pi^- l^+ \nu_l $ and $ f_2 = \pi^+ l^- \bar{\nu}_l $,
$ \left| A( \dt; \pi^- l^+ \nu_l, \pi^+ l^- \bar{\nu}_l ) \right|^2 $
is an even function of $ \dt $ if $ CPT $ is not violated.
 Thus the following characterizes $ CPT $ violation
\begin{eqnarray}
 \lefteqn{ {\cal A}_{ l^+ l^- } } \nonumber \\
 & \equiv &
\displaystyle{
 \frac{
  \int_0^\infty d(\dt)
      \left| A( \dt; \pi^- l^+ \nu_l, \pi^+ l^- \bar{\nu}_l )
      \right|^2
 - \int_{-\infty}^0 d(\dt)
 \left| A( \dt; \pi^- l^+ \nu_l, \pi^+ l^- \bar{\nu}_l ) \right|^2
 }
 {
  \int_0^\infty d(\dt)
      \left| A( \dt; \pi^- l^+ \nu_l, \pi^+ l^- \bar{\nu}_l )
      \right|^2
 + \int_{-\infty}^0 d(\dt)
 \left| A( \dt; \pi^- l^+ \nu_l, \pi^+ l^- \bar{\nu}_l ) \right|^2
 }
} \nonumber \\
 & = & - 4 \Rp \BD.
\label{ eqn:cll }
\end{eqnarray}
 Here $ \BD $ is defined by
\begin{equation}
 2 \BD \equiv 2 \Delta +  x_l -  \bar{x}_l^*.
\end{equation}
\par
 The relative time probability distribution for
$ f_1 = \pi^+ l^- \bar{\nu}_l, f_2 = \pi^+ l^- \bar{\nu}_l $
and the one for $ f_1 = \pi^- l^+ \nu_l, f_2 = \pi^- l^+ \nu_l $
differ only by the overall magnitude,
so that the following quantity seems useful
\begin{eqnarray}
 \lefteqn{ {\cal A}(\pi^+ l^- \bar{\nu}_l, \pi^- l^+ \nu_l ) }
\nonumber \\
 & \equiv &
\displaystyle{
 \frac{
  \int_{ -\infty }^{\infty} d( \dt )
    \left| A( \dt; \pi^+ l^- \bar{\nu}_l, \pi^+ l^- \bar{\nu}_l )
    \right|^2
 - \int_{ -\infty }^{\infty} d( \dt )
     \left| A( \dt; \pi^- l^+ \nu_l, \pi^- l^+ \nu_l ) \right|^2
 }
 {
  \int_{ -\infty }^{\infty} d( \dt )
    \left| A( \dt; \pi^+ l^- \bar{\nu}_l, \pi^+ l^- \bar{\nu}_l )
    \right|^2
 + \int_{ -\infty }^{\infty} d( \dt )
     \left| A( \dt; \pi^- l^+ \nu_l, \pi^- l^+ \nu_l ) \right|^2
 }
}
\nonumber \\
 & = & -4 \Rp ( \epsilon_0 - y_l ).
\label{ eqn:cplpl }
\end{eqnarray}
\par
 The $ \pi \pi $ $ ( \  = \pi^+ \pi^- $ or $ \pi^0 \pi^0 ) $ channel
can be used to measure $ \delta_l (\infty) $ ( see Eq.
$ \left( \ref{ eqn:Ksas } \right) $ ) through the following quantity
\begin{eqnarray}
 \lefteqn{ {\cal A}_{\pi\pi,l^+ l^-}^+ } \nonumber \\
 & \equiv &
\displaystyle{
 \frac{
  \int_0^{\infty} d(\dt)
    \left| A( \dt; \pi^- l^+ \nu_l, \pi\pi ) \right|^2
    - \int_0^{\infty} d(\dt)
        \left| A( \dt; \pi^+ l^- \bar{\nu}_l, \pi\pi )
        \right|^2
 }
 {
  \int_0^{\infty} d(\dt)
     \left| A( \dt; \pi^- l^+ \nu_l, \pi\pi ) \right|^2
 + \int_0^{\infty} d(\dt)
     \left| A( \dt; \pi^+ l^- \bar{\nu}_l, \pi\pi )
     \right|^2
 }
}
\nonumber \\
 & = &
   2 \Bigl\{ \Rp( \epsilon_0 - y_l ) - \Rp( \BD )
   - 2 \left| \eta_{\pi\pi} \right|
\displaystyle{
 \frac{\wl}{\ws}
}
\cos \phi_{SW}  \cdot \cos( \phi_{\pi\pi} + \phi_{SW} ) \Bigl\}.
\label{ eqn:cpppll }
\end{eqnarray}
 Numerically $ \wl / \ws, \, \left| \eta_{\pi\pi} \right| \,
\sim 10^{-3} $
so that $ \left( \ref{ eqn:cpppll } \right) $
reduces to the usual asymmetric quantity $ \delta_l ( \infty ) $;
\begin{eqnarray}
 \delta_l ( \infty ) & \equiv &
     \frac{ \Gamma_L( \pi^- l^+ \nu_l )
        - \Gamma_L( \pi^+ l^- \bar{\nu}_l ) }
     { \Gamma_L( \pi^- l^+ \nu_l )
        + \Gamma_L( \pi^+ l^- \bar{\nu}_l ) }
\nonumber \\
 & = &
2 \{ \Rp( \epsilon_0 - y_l ) - \Rp( \BD ) \}.
\label{ eqn:Ksas }
\end{eqnarray}
 We can construct the quantity which is the counterpart of
$ {\cal A}_{ \pi \pi, \, l^+ l^- }^+ $ as follows;
\begin{eqnarray}
 \lefteqn{ {\cal A}_{\pi\pi,l^+ l^- }^- } \nonumber \\
 & \equiv &
\displaystyle{
 \frac{
  \int_{-\infty}^0 d(\dt) \left| A( \dt; \pi^- l^+ \nu_l, \pi\pi )
                          \right|^2
 - \int_{-\infty}^0 d(\dt)
         \left| A( \dt; \pi^+ l^- \bar{\nu}_l, \pi\pi ) \right|^2
 }
 {
  \int_{-\infty}^0 d(\dt)
        \left| A( \dt; \pi^- l^+ \nu_l, \pi\pi ) \right|^2
 + \int_{-\infty}^0 d(\dt)
        \left| A( \dt; \pi^+ l^- \bar{\nu}_l, \pi\pi )
        \right|^2
 }
}
\nonumber \\
 & = &
2 \Bigl\{ \Rp( \epsilon_0 - y_l ) - \Rp( \BD )
    - 2 \left| \eta_{\pi\pi} \right|
\cos \phi_{SW} \cdot \cos( \phi_{\pi\pi} - \phi_{SW} ) \Bigl\}.
\label{ eqn:cmppll }
\end{eqnarray}
 The value of $ \phi_{\pi \pi} $
will be fixed by the measurement of $ {\cal A}^-_{\pi\pi,l^+l^-} $
\cite{Buchanan} after determination of $ \delta_l(\infty) $ and,
$ \left| \eta_{\pi\pi} \right| $, for example,
by using $ {\cal A}^+_{\pi\pi,l^+l^-} $,
and $ \left| A( \pi\pi,t ) \right|^2 $
( see Eq.$ \left( \ref{ eqn:inpipi } \right) $ in
Sec.\ref{ subsec:tas } ).
%
%
%%%%%%%%%%%%%%%%%%%%%%%%%%%%%%%%%%%%%%%%%%%%%%%%%%%%%%%%%%%%%%
%                                                            %
%             Inclusive semileptonic decay mode              %
%                                                            %
%%%%%%%%%%%%%%%%%%%%%%%%%%%%%%%%%%%%%%%%%%%%%%%%%%%%%%%%%%%%%%
%
%
\par
 Inclusive semileptonic decay mode,
in which one final state is specified
to be $ \pi^+ l^- \bar{\nu}_l $ or $ \pi^- l^+ \nu_l $
and the other one is not restricted,
may enable us to perform more accurate measurements of parameters
through accumulating larger number of events.
 $ {\cal A}^{\rm inclusive}_{ l^+ l^- } $ is defined by
\begin{eqnarray}
 {\cal A}^{\rm inclusive}_{ l^+ l^- } & \equiv &
\displaystyle{
 \frac{ \int_0^\infty dt \left| A( \pi^- l^+ \nu_l, t ) \right|^2
  - \int_0^\infty dt \left| A( \pi^+ l^- \bar{\nu}_l, t ) \right|^2
 }
 {
  \int_0^\infty dt \left| A( \pi^- l^+ \nu_l, t ) \right|^2
  + \int_0^\infty dt \left| A( \pi^+ l^- \bar{\nu}_l, t ) \right|^2
 }
}
\nonumber \\
& = & 2 \{ \Rp( \epsilon_0 - y_l ) - \Rp( \BD ) \},
\label{ eqn:cinll }
\end{eqnarray}
 with
\begin{equation}
 \left| A( \pi^- l^+ \nu_l, t ) \right|^2
\equiv \sum_{f} \int_0^\infty dt^\prime
\left| A( \pi^- l^+ \nu_l,t; \, f,t^\prime ) \right|^2.
\end{equation}
 So $ {\cal A}^{\rm inclusive}_{ l^+ l^- } $ is equal
to $ \delta_l (\infty) $
apart from numerically negligible constants.
\par
 The counterpart to $ \delta_l(\infty) $, $ \bar{\delta}_l(\infty) $
corresponding to the $ K_S $ decay can be determined from
\begin{eqnarray}
 \bar{\delta}_l ( \infty ) & \equiv &
\frac{ \Gamma_S ( \pi^- l^+ \nu_l )
- \Gamma_S ( \pi^+ l^- \bar{\nu}_l ) }
{ \Gamma_S ( \pi^- l^+ \nu_l ) + \Gamma_S ( \pi^+ l^- \bar{\nu}_l ) }
\nonumber \\
 & = &
2 \{ \Rp( \epsilon_0 - y_l ) + \Rp( \BD ) \}.
\label{ eqn:Kbas }
\end{eqnarray}
 Its observation is restricted to the semileptonic $ K_S $
decay measurement with $ K_L $ tagging.
 Thus it appears only as the coefficients of $ e^{ - \ws \dt } $
( or $ e^{ -\ws t } $ ) in the expressions of
the various relative time distribution functions
( or the time distributions of semileptonic decay ).
%
%
%%%%%%%%%%%%%%%%%%%%%%%%%%%%%%%%%%%%%%%%%%%%%%%%%%%%%%%%%%%%%%%%
%                                                              %
%                         Subsection 3.2                       %
%                                                              %
%%%%%%%%%%%%%%%%%%%%%%%%%%%%%%%%%%%%%%%%%%%%%%%%%%%%%%%%%%%%%%%%
%
%
\subsection{ $ \dt $ dependent asymmetries }
\label{ subsec:tas }
 With recent development in technology,
it is now possible to increase the resolution of
the decay vertex point determination to about $ 20 {\mu m} $.
 This allows us to obtain additional information.
 More precisely, some interference terms whose time dependences
are periodic with exponential damping,
give large contributions at $ t = {\cal O}( \taus ) $,
where $ \taus = 1/\ws $.
 The effects of such interference terms can be better observed
in asymmetries which depend on $ \dt $.
\par
 To make it more explicit, we list the following quantity
\begin{eqnarray}
 \lefteqn{ {\cal A}^{l^+ l^-}( \dt ) }
\nonumber \\
 & \equiv &
\displaystyle{
 \frac{
  \left| A( \dt>0; \pi^- l^+ \nu_l, \pi^+ l^- \bar{\nu}_l ) \right|^2
 - \left| A( -\dt; \pi^- l^+ \nu_l, \pi^+ l^- \bar{\nu}_l ) \right|^2
 }
 {
  \left| A( \dt>0; \pi^- l^+ \nu_l, \pi^+ l^- \bar{\nu}_l ) \right|^2
 + \left| A( -\dt; \pi^- l^+ \nu_l, \pi^+ l^- \bar{\nu}_l ) \right|^2
 }
}
\nonumber \\
 & = & -
\displaystyle{
 \frac{
  4 \Rp (\BD) ( e^{ -\wl\dt } - e^{ -\ws\dt } )
 + 8 \Ip (\BD) e^{ -\frac{\wt}{2} \dt } \sin( \dm\dt )
 }
 {
  e^{ -\wl\dt } + e^{ -\ws\dt } + 2 e^{ -\frac{\wt}{2} \dt }
    \cos( \dm\dt )
 }
},
\label{ eqn:tll }
\end{eqnarray}
 which will reduce to the result obtained by Ref.\cite{Buchanan}
when $ \DSDQ $ violating parameter $ x_l, \bar{x}_l $
is set equal to 0.
 When $ \dt = ( 0 \sim 15 ) \taus $,
the contribution of sin-term to the total behavior
is still large due to the factor $ \frac{1}{2} $ in the exponential
so that we can make a best fit to deduce the value of $ \Ip \BD $.
 Note that $ \dm\dt $ varies by a factor of $ 2 \pi $
for the range $ 0 \alt \dt \alt 15 \taus $.
 For large $ \dt ( \dt \agt 20 \taus ) $, $ {\cal A}^{l^+ l^-} $
is sensitive to the term proportional to $ \Rp \BD $.
 Therefore, both $ \Rp \BD $ and $ \Ip \BD $ can be observed
by examining the time evolution during $ \dt = ( 0 \sim 15 ) \taus $
and also $ \dt \agt 20 \taus $.
\par
 From the $ f_1 = \pi^+ \pi^-,\  f_2 = \pi^0 \pi^0 $ mode,
the value of $ \Ip \epsilon^\prime / \epsilon $ can be measured
by observing the small $ \dt $ behavior
( $ \dt = ( 0 \sim 15 )/ \taus $ ) of the quantity \cite{Buchanan}
\begin{eqnarray}
 \lefteqn{ {\cal A} ( \dt; \pi^+ \pi^-, \pi^0 \pi^0 ) }
\nonumber \\
 & \equiv &
\displaystyle{
 \frac{
  \left| A( \dt>0; \pi^+ \pi^-, \pi^0 \pi^0 ) \right|^2
 - \left| A( -\dt; \pi^+ \pi^-, \pi^0 \pi^0 ) \right|^2
 }
 { \left| A( \dt>0; \pi^+ \pi^-, \pi^0 \pi^0 ) \right|^2
 + \left| A( -\dt; \pi^+ \pi^-, \pi^0 \pi^0 ) \right|^2
 }
}
\nonumber \\
 & = & 3
\displaystyle{
 \frac{
  \Rp ( \epsilon^\prime /\epsilon )
     ( e^{ -\wl \dt } - e^{ -\ws\dt } )
 - 2 e^{ -\frac{\wt}{2} \dt } \Ip ( \epsilon^\prime /\epsilon )
        \sin( \dm\dt )
 }
 { e^{ -\wl\dt } + e^{ -\ws\dt } - 2 e^{ -\frac{\wt}{2} \dt }
        \cos( \dm\dt )
 }
}.
\label{ eqn:tpipi }
\end{eqnarray}
%
%
%%%%%%%%%%%%%%%%%%%%%%%%%%%%%%%%%%%%%%%%%%%%%%%%%%%%%%%%%%%%%
%                                                           %
%            Inclusive \pi\pi time distribution             %
%                                                           %
%%%%%%%%%%%%%%%%%%%%%%%%%%%%%%%%%%%%%%%%%%%%%%%%%%%%%%%%%%%%%
%
%
\par
 From the inclusive $ \pi\pi $ decay mode, we cannot construct
any asymmetric quantities.
 However the conventional method of detecting
$ \left| \eta_{\pi\pi} \right| $
is also applicable at a $ \phi $ factory \cite{Buchanan}
\begin{eqnarray}
 \left| A( \pi\pi, t ) \right|^2 & \equiv &
\sum_f \int_0^\infty dt^\prime \left| A( \pi\pi,t; f,t^\prime )
\right|^2     \nonumber \\
 & = &
\displaystyle{
 \frac{1}{2} \left| \left< \pi\pi |T| K_S \right> \right|^2 \Bigl\{
\left| \eta_{\pi\pi} \right|^2 e^{ -\wl t } + e^{ -\ws t }
}
\nonumber \\
 & & -
2 \left| \eta_{\pi\pi} \right| \left| \left< K_S | K_L \right>
  \right|
e^{ -\frac{\wt}{2} t } \cos( \dm t - \phi_{\pi\pi} - \phi_{LS} )
\Bigl\}.
\label{ eqn:inpipi }
\end{eqnarray}
 where
\begin{eqnarray}
 \left< K_L | K_S \right> & \equiv &
\left| \left< K_L | K_S \right> \right| e^{ i \phi_{LS} }
\nonumber \\
 & = & 2 \{ \Rp \epsilon_0 + i \Ip \Delta \}.
\label{ eqn:innerproduct }
\end{eqnarray}
\par
 As noted above in Sec.3.1,
the time evolutions of
$ \left| A( \dt; \pi^+ l^- \bar{\nu}_l, \pi^+ l^- \bar{\nu}_l )
  \right|^2 $
and $ \left| A( \dt; \pi^- l^+ \nu_l, \pi^- l^+ \nu_l ) \right|^2 $
are the same so that the quantity
\begin{eqnarray}
 \lefteqn{ {\cal A}( \dt; \pi^+ l^- \bar{\nu}_l, \pi^- l^+ \nu_l ) }
\nonumber \\
 & \equiv &
\displaystyle{
 \frac{
  \left| A( \dt; \pi^+ l^- \bar{\nu}_l, \pi^+ l^- \bar{\nu}_l )
  \right|^2
  - \left| A( \dt; \pi^- l^+ \nu_l, \pi^- l^+ \nu_l ) \right|^2
 }
 {
  \left| A( \dt; \pi^+ l^- \bar{\nu}_l, \pi^+ l^- \bar{\nu}_l )
  \right|^2
  + \left| A( \dt; \pi^- l^+ \nu_l, \pi^- l^+ \nu_l ) \right|^2
 }
}
\nonumber \\
 & = &
- 4 \Rp ( \epsilon_0 - y_l ),
\label{ eqn:tplpl }
\end{eqnarray}
 is constant in $ \dt $ \cite{KEK}.
 It appears to be the most suitable observable for determining
$ \Rp( \epsilon_0 - y_l ) $ since it is possible
to take the average of
the values of
$ {\cal A} ( \dt; \pi^+ l^- \bar{\nu}_l, \pi^- l^+ \nu_l ) $
at various relative times $ \dt $.
\par
 Corresponding to the quantities $ {\cal A}^+_{ \pi\pi, l^+ l^- } $
and $ {\cal A}^-_{ \pi\pi, l^+ l^- } $,
 one can consider the following respective asymmetries
\begin{eqnarray}
 \lefteqn{
  {\cal A}^{\pi\pi}
    ( \dt>0; \pi^- l^+ \nu_l, \pi^+ l^- \bar{\nu}_l ) }
\nonumber \\
 & \equiv &
\displaystyle{
 \frac{
  \left| A( \dt>0; \pi^- l^+ \nu_l, \pi\pi ) \right|^2
 - \left| A( \dt>0; \pi^+ l^- \bar{\nu}_l, \pi\pi ) \right|^2
 }
 {
  \left| A( \dt>0; \pi^- l^+ \nu_l, \pi\pi ) \right|^2
 + \left| A( \dt>0; \pi^+ l^- \bar{\nu}_l, \pi\pi ) \right|^2
 }
}
\nonumber \\
 & = & 2
\displaystyle{
 \frac{
  ( \Rp( \epsilon_0 - y_l ) - \Rp( \BD ) ) e^{ -\wl\dt }
 - \left| \eta_{\pi\pi} \right| e^{ -\frac{\wt}{2} \dt }
 \cos( \dm\dt + \phi_{\pi\pi} )
 }
 {
  e^{ -\wl\dt } + \left| \eta_{\pi\pi} \right|^2 e^{ -\ws\dt }
 }
},
\nonumber \\
\label{ eqn:tpppll }
\end{eqnarray}
\begin{eqnarray}
 \lefteqn{ {\cal A}^{\pi\pi}
     ( \dt<0; \pi^- l^+ \nu_l, \pi^+ l^- \bar{\nu}_l ) }
\nonumber \\
 & \equiv &
\displaystyle{
 \frac{
  \left| A( \dt<0; \pi^- l^+ \nu_l, \pi\pi ) \right|^2
 - \left| A( \dt<0; \pi^+ l^- \bar{\nu}_l, \pi\pi ) \right|^2
 }
 {
  \left| A( \dt<0; \pi^- l^+ \nu_l, \pi\pi ) \right|^2
 + \left| A( \dt<0; \pi^+ l^- \bar{\nu}_l, \pi\pi ) \right|^2
 }
}
\nonumber \\
 & = & 2
\displaystyle{
 \frac{
  ( \Rp( \epsilon_0 - y_l ) - \Rp( \BD ) ) e^{ -\ws\adt }
 - \left| \eta_{\pi\pi} \right| e^{ -\frac{\wt}{2} \adt }
 \cos( \dm\adt - \phi_{\pi\pi} )
 }
 {
 e^{ -\ws\adt } + \left| \eta_{\pi\pi} \right|^2 e^{ -\wl\adt }
 }
}.
\nonumber \\
\label{ eqn:tppmll }
\end{eqnarray}
 From the large $ \dt $ behavior of
$ {\cal A}^{\pi\pi}
     ( \dt>0; \pi^- l^+ \nu_l, \pi^+ l^- \bar{\nu}_l ) $,
the combination $ \left( \Rp( \epsilon_0 - y_l ) - \Rp \BD \right) $
which is just half of $ \delta_l (\infty) $ will be measured.
 The interference term in
$ {\cal A}^{\pi\pi}
 ( \dt<0; \pi^- l^+ \nu_l, \pi^+ l^- \bar{\nu}_l ) $
may provide us with the value of $ \phi_{\pi\pi} $.
%
%
%%%%%%%%%%%%%%%%%%%%%%%%%%%%%%%%%%%%%%%%%%%%%%%%%%%%%%%%%%%%%%
%                                                            %
%             Inclusive semileptonic decay mode              %
%                                                            %
%%%%%%%%%%%%%%%%%%%%%%%%%%%%%%%%%%%%%%%%%%%%%%%%%%%%%%%%%%%%%%
%
%
\par
 Finally, in the semileptonic inclusive decay mode one can construct
the following asymmetric quantity
\begin{eqnarray}
 {\cal A}^{\rm inclusive}_{ l^+ l^- } (t) & \equiv &
\displaystyle{
 \frac{
  \left| A( \pi^- l^+ \nu_l, t ) \right|^2
 - \left| A( \pi^+ l^- \bar{\nu}_l, t ) \right|^2
 }
 {
  \left| A( \pi^- l^+ \nu_l, t ) \right|^2
 + \left| A( \pi^+ l^- \bar{\nu}_l, t ) \right|^2
 }
}
\nonumber \\
 & = &
\displaystyle{ \frac{2}{ e^{ -\wl t} + e^{ -\ws t} } }
\Bigl[ \left\{ \Rp( \epsilon_0 - y_l ) - \Rp( \BD ) \right\}
    e^{ -\wl t }
\nonumber \\
 & \quad & + \left\{ \Rp( \epsilon_0 - y_l ) + \Rp( \BD ) \right\}
    e^{ -\ws t }
\nonumber \\
 & \quad & - \left| \left< K_S | K_L \right> \right|
    e^{ -\frac{\wt}{2} t } \cos( \dm t - \phi_{LS} )
\Bigl],
\label{ eqn:tinll }
\end{eqnarray}
 Hence, as can be seen
in Eq.$ \left( \ref{ eqn:innerproduct } \right) $,
we can determine the values of $ \Rp \epsilon_0 $ and $ \Ip \Delta $
from the behavior of $ {\cal A}^{\rm inclusive}_{ l^+ l^- } (t) $
around $ \dt = (0 \sim 15) \taus $.
\par
 We summarize the measurement at a $ \phi $ factory in Table
\ref{ tab:phi }.
 Measurements of the parameters from $ \ws $ to $ \epsilon $
are the same as in Ref.\cite{Buchanan}.
 We include their results in that table for comleteness.
 Note that $ \Rp \Delta,\ \Rp x_l $ and $ \Rp \bar{x}_l $ appear
only as a combination of
$ 2 \BD = 2 \Delta + ( x_l - \bar{x}_l^* ) $.
 Thus we {\it cannot} measure $ \Rp \Delta $ and
$ \Rp( x_l - \bar{x}_l ) $
{\it separately} at a $ \phi $ factory \cite{Patera}.
 To make further progress, we must turn to the experiments
with tagged kaon beam.
%
%
%%%%%%%%%%%%%%%%%%%%%%%%%%%%%%%%%%%%%%%%%%%%%%%%%%%%%%%%%%%%%%%%%%
%                                                                %
%                             Section 4                          %
%                                                                %
%%%%%%%%%%%%%%%%%%%%%%%%%%%%%%%%%%%%%%%%%%%%%%%%%%%%%%%%%%%%%%%%%%
%
%
\section{ Further measurements with tagged $ K $ }
\label{ sec:lear }
 We now ask if all the parameters remained undetermined
at the $ \phi $ factory
can be fixed by the experiments with kaon tagging.
\par
 We can use the probabilities $ \PA, \BA, \PL $ and $ \BL $ in Sec.2
to define time dependent asymmetries.
 The most familiar one is $ \beta(t) $ which have been often referred
\cite{Tanner,Kabir} in the context of Kabir's direct test
for $ T $ violation.
\begin{eqnarray}
  \beta_l (t) & \equiv &
 \displaystyle{ \frac{ \BA - \PL }{ \BA + \PL } }
 \nonumber \\
& = &
\displaystyle{
 \frac{1}{ e^{ -\wl t } + e^{ -\ws t } - 2 e^{ -\frac{\wt}{2} t }
    \cos( \dm t ) }
\Bigl[
A e^{ -\wl t } + B e^{ -\ws t } + C e^{ -\frac{\wt}{2} t }
    \cos( \dm t )
}
\nonumber \\
 & & + D e^{ -\frac{\wt}{2} t } \sin( \dm t ) \Bigl ].
\label{ eqn:asybeta }
\end{eqnarray}
 The explicit expressions for the coefficients $ A-D $ are shown
in Table \ref{ tab:abcoeff }.
 The size of $ \beta_l(t) $ is of order of small parameters;
 it is zero as long as $ CP, CPT $ symmetry and $ \DSDQ $ rule hold.
 We define a parameter $ \alpha_l(t) $ which has a similar property:
\begin{eqnarray}
 \alpha_l (t) & \equiv &
\displaystyle{ \frac{ \BL - \PA }{ \BL + \PA } }
\nonumber \\
 & = &
\displaystyle{
 \frac{1}{ e^{ -\wl t } + e^{ -\ws t }
 + 2 e^{ -\frac{\wt}{2} t } \cos( \dm t ) }
 \Bigl[
 A e^{ -\wl t } + B e^{ -\ws t } + C e^{ -\frac{\wt}{2} t }
     \cos( \dm t )
}
\nonumber \\
 & & + D e^{ -\frac{\wt}{2} t } \sin( \dm t ) \Bigl].
\label{ eqn:asyalpha }
\end{eqnarray}
 The coefficients $ A-D $ are also listed
in Table \ref{ tab:abcoeff }.
\par
 If it is possible to achieve the time resolution much
less than $ \taus $,
all the coefficients $ A,B,C $ and $ D $ can be obtained
in $ \alpha_l(t) $ and $ \beta_l(t) $.
 We can then determine $ \Rp \Delta and \Rp( x_l - \bar{x}_l ) $
as shown in Table \ref{ tab:abcoeff }.
 However $ \beta_l(t) $, and $ \alpha_l(t) $ are not enough
to find the values of $ \Rp x_l $ and $ \Rp \bar{x}_l $.
 The values of $ \Ip x_l $ and $ \Ip \bar{x}_l $ also remain
undetermined.
 To remedy this situation we construct 4 further asymmetries
as follows;
\begin{eqnarray}
 \delta_l (t) & \equiv &
\displaystyle{ \frac{ \PA - \PL }{ \PA + \PL } },
\quad \quad
 \bar{ \delta }_l (t) \equiv
\displaystyle{ \frac{ \BA - \BL }{ \BA + \BL } }, \nonumber \\
 \gamma_l (t) & \equiv &
\displaystyle{ \frac{ \BA - \PA }{ \BA + \PA } },
\quad \quad
 \bar{ \gamma }_l (t) \equiv
\displaystyle{ \frac{ \BL - \PL }{ \BL + \PL } },
\label{ eqn:asy }
\end{eqnarray}
 Explicit expressions for these asymmetries have the following
common form:
\begin{eqnarray}
 \Lambda & = & \displaystyle{ \frac{2}{ e^{ -\wl t } +
       e^{ -\ws t } } } \Bigl[
A e^{ -\wl t } + B e^{ -\ws t } + C e^{ -\frac{\wt}{2} t }
    \cos( \dm t )
+ D e^{ -\frac{\wt}{2} t } \sin( \dm t ) \nonumber \\
 & & +
\displaystyle{
 E \frac{ e^{ -\wl t } }{ e^{ -\wl t } + e^{ -\ws t } }
   e^{ -\frac{\wt}{2} t } \cos( \dm t )
 + F \frac{ e^{ -\ws t } }{ e^{ -\wl t } + e^{ -\ws t } }
   e^{ -\frac{\wt}{2} t } \cos( \dm t )
}
\nonumber \\
 & & +
\displaystyle{
 G \frac{ e^{ -\wt t } }{ e^{ -\wl t } + e^{ -\ws t } }
   \cos^2 ( \dm t )
+ H \frac{ e^{ -\wt t } }{ e^{ -\wl t } + e^{ -\ws t } }
   \sin( \dm t ) \cos( \dm t )
}
\Bigl].
\label{ eqn:Lambda }
\end{eqnarray}
 The coefficients in each $ \Lambda $ are shown
in Table \ref{ tab:lambdacoeff }.
\par
 Note that the difference between the functional form
of $ \Lambda $ and those of $ \alpha_l $ and $ \beta_l $
originates from the fact that
all $ \Lambda^\prime $ s approach $ 1 $ in the limit of
$ CP, CPT $ conservation and exact $ \DSDQ $ rule.
 We also point out that $ \Lambda^\prime $s may be useful
as they do not depend on an overall normalization factor.
\par
 We summarize the procedure for determining $ x_l, \bar{x}_l $
from $ \Lambda $'s in Table \ref{ tab:lambdaexp }.
 Suppose that the experimental analysis can make a best fit
to the time dependent data using the functional form given above
with $ A-H $.
 Then we can determine all the parameters
as shown in Table \ref{ tab:lambdaexp }.
%
%
%
%
%%%%%%%%%%%%%%%%%%%%%%%%%%%%%%%%%%%%%%%%%%%%%%%%%%%%%%%%%%%%%%
%                                                            %
%                         Section 5                          %
%                                                            %
%%%%%%%%%%%%%%%%%%%%%%%%%%%%%%%%%%%%%%%%%%%%%%%%%%%%%%%%%%%%%%
%
%
\section{ Discussion and conclusion }
\label{ sec:discon }
 The main aim of high energy physics, at present,
is to find the physics beyond the standard model.
 One possible approach is to build larger and larger accelerators
which may be capable of producing new particles.
 Another way is to search for fine deviation from the standard model
predictions which can be attributed to quantum effects
of new physics.
 The most effective way to proceed with the latter, in our opinion,
is to search for deviations from the standard model predictions on
$ CP $ violating observables.
 The standard model does not offer any understanding
as to the origin of the $ CP $ violation.
 If new physics is to show up at all,
it is most likely to appear in $ CP $ violating observables.
\par
 Our aim of this paper is to establish a systematic method for
testing the fundamendal symmetries, $ CP,T, $ and $ CPT $,
in the future collider experiments.
 The most attractive candidate for that purpose is
$ \phi $ factory project.
 We also take into account of the possibility of large $ \DSDQ $ rule
violation in the $ K^0 $ - $ \bar{K}^0 $ system,
which may be recognized as a result of some high energy physics
beyond the standard model.
\par
 In order to measure $ CP, T, CPT, $ and $ \DSDQ $ rule violations,
we construct all possible asymmetries,
accessible to the neutral kaon system,
and evaluate them in terms of various parameters characterizing
the symmetry violations.
 We summarize our findings as follows;
\begin{enumerate}
\item
 As summarized in Table \ref{ tab:phi }, we showed that
not all parameters can be determined
if we focus our attention only on the $ \phi $ factory experiment
as shown in Sec.\ref{ sec:phi }.
\item
 With respect to the $ CPT $ violation in the $ K^0 $-$\bar{K}^0 $
mass matrix we see there that only the combination
$ \Rp \BD = \Rp \Delta + \Rp(x_l - \bar{x}_l) $ can be determined,
and that the seperate measurement of $ \Rp \Delta $ and
$ \Rp( x_l + \bar{x}_l ) $ cannot be achieved.
\item
 With respect to $ CPT $ violation in the amplitudes,
the parameters $ (1-\Rp{A_I}/\Rp\bar{A_I})\ (I=0,2) $ associated
with $ 2\pi $ decay mode also remain undertermined.
 This is because
from Eq.$ \left( \ref{ eqn:epsilonprime } \right) $
these are given by
\begin{eqnarray}
 \displaystyle{
  1 - \frac{\Rp\bar{A_0}}{\Rp{A_0}}
 } &=&
 \displaystyle{
  2 \left( \Rp\epsilon - \Rp\epsilon_0 + \Rp\Delta \right)
 }, \nonumber \\
 \displaystyle{
  1 - \frac{\Rp\bar{A_2}}{\Rp{A_2}}
 }
 &=&
 \displaystyle{
  \Rp\left( \frac{2\sqrt{2} \epsilon^\prime }{ \omega }
              e^{ -i(\delta_2 - \delta_0 ) }
      \right)
  + \left( 1 - \frac{\Rp\bar{A_0}}{\Rp{A_0}} \right)
 },
\nonumber
\end{eqnarray}
 and $ \Rp\Delta $ is not determined.
\item
 However, as shown in Sec.\ref{ sec:lear },
this lack of information can be improved by appealing
to the observation of $ \beta_l(t) $ and $ \alpha_l(t) $
in the experiments
which can provide $ K^0 $ and $ \bar{K}^0 $ beams seperately.
 Then, as promised, 3 further $ CPT $ violating parameters
$ \Rp\Delta, \Rp( x_l - \bar{x}_l ) $
and $ (1-\Rp\bar{A_I}/\Rp{A_I}) (I=0,2) $ can be determined.
\item
 Determination of the values of $ \DSDQ $ rule violating parameters
$ x $ and $ \bar{x}_l $ requires the measurement of asymmetries
of the kind that are order 1,
in contrast to the previous ones which reduce to 0
in the exact symmetry limit.
 These asymmetries $ \delta_l(t), \bar{\delta}_l(t), \gamma_l(t) $
and $ \bar{\gamma}_l(t) $ have the common functional dependence
on time $ t $, denoted as $ \Lambda $ in Sec.\ref{ sec:lear }.
 $ \Lambda $ consists of 8 parameters
which are to be adjusted from the experiments.
\end{enumerate}
\par
 In Table \ref{ tab:phi },
we have included the procedure for determining
the $ CP $ and $ T $ violating parameters,
$ ( \Ip{A_I}/\Rp{A_I} - \Ip{\bar{A}_I}/\Rp{A_I} ) ( I=0,2 )$
in the $ 2 \pi $ decay amplitudes,
as possibly measured at a $ \phi $ factory.
 This is true only if we {\it assume} that $ \Ip\epsilon_0 = 0 $
since $ \Ip\epsilon_0 $ as well as $ \Ip{y_l} $ cannot appear
in any observables in the experiments,
so they cannot be measured in any way.
 In fact from Eq.$ \left( \ref{ eqn:epsilonprime } \right) $
we then have
\begin{equation}
 \displaystyle{
  \frac{ \Ip{A_0} }{ \Rp{A_0} } - \frac{ \Ip\bar{A}_0 }{ \Rp{A_0} }
  = 2 \left( \Ip\epsilon + \Ip\Delta \right)
 },
\end{equation}
\begin{eqnarray}
 \displaystyle{
  \frac{ \Ip{A_2} }{ \Rp{A_2} } - \frac{ \Ip\bar{A}_2 }{ \Rp{A_2} }
 }
&=&
\displaystyle{
 \Ip\left(
          \frac{2\sqrt{2} \epsilon^\prime}{\omega}
                e^{-i(\delta_2-\delta_0)}
    \right)
} \nonumber \\
 &+&
 \displaystyle{
  \left(
     \frac{ \Ip{A_0} }{ \Rp{A_0} }
       - \frac{ \Ip\bar{A}_0 }{ \Rp{A_0} }
  \right)
 }.
\end{eqnarray}
 The fact that $ \Ip\epsilon_0 $ and $ \Ip y_l $ does not appear
in the measurable quantities
in the experiments with seperate $ K^0 $ and $ \bar{K}^0 $ beams
has already been mentioned by Tanner and Dalitz \cite{Tanner}.
 We see that we are confronted with the same obstracle
in the $ \phi $ factory experiments.
 The reason is that they always appear accompanying 1 so that
they drop out when we take the absulute square of amplitudes
to get probabilities to the first order
with respect to small parameters \cite{Tanner}.
 If one persists to retain $ \Ip\epsilon_0 $,
the combination
\begin{equation}
 \frac{1}{2}
 \left(
       \frac{\Ip{A_0}}{\Rp{A_0}} - \frac{\Ip\bar{A}_0}{\Rp{A_0}}
 \right) + \Ip\epsilon_0 = \Ip\epsilon + \Ip\Delta
\end{equation}
is determined, but $ ( \Ip{A_2}/\Rp{A_2} - \Ip\bar{A}_2/\Rp{A_2} ) $
remains undetermined.
 To seperate the
$ \Ip{A_0}/\Rp{A_2} - \Ip{\bar{A}_2}/\Rp{\bar{A}_2} $
from $ \Ip\epsilon_0 $,
we must appeal to the experiment that will reach to the order
$ \epsilon^2 \simeq 10^{-6} $ precision.
\par
 We refrain from making estimates of errors on various measurements
as they depend too much on the actual experimental configurations.
 We, however, make a few comments along this line based on
an elabolate work by Buchanan {\it et.al.}\cite{Buchanan}
with respect to a $ \phi $ factory.
 They performed a quantitative analysis
under the experimental situation at a $ \phi $ factory.
 Considering the possibility of $ \DSDQ $ rule violation,
we can still apply their result with some modifications.
\par
 According to their results, $ \delta_l(\infty) $ and
$ \bar{\delta}_l(\infty) $
is measured to the precision $ \pm 0.013 \times 10^{-3} $ and
$ \pm 0.13 \times 10^{-3} $ respectively.
 From the relation
\begin{equation}
 \Rp\BD = - \frac{1}{4} \left(
                           \delta_l(\infty) - \bar{\delta}_l(\infty)
                        \right),
\end{equation}
we can determine $ \Rp\BD $, not $ \Rp\Delta $, within the error
$ \pm 0.032 \times 10^{-3} $.
 Also we have
\begin{equation}
 \Rp( \epsilon_0 - y_l ) =
\frac{1}{4} \left(
               \delta_l(\infty) + \bar{\delta}_l(\infty)
            \right).
\end{equation}
 Hence we can obtain the value of $ \Rp( \epsilon_0 - y_l ) $
with the accuracy $ \pm 0.032 \times 10^{-3} $.
 The observation of inclusive semileptonic decay channels gives
$ \Rp \epsilon_0 $ to $ \pm 0.18 \times 10^{-3} $ \cite{Buchanan}.
 So, when combining this with the value of
$ \Rp( \epsilon_0 - y_l ) $,
$ CPT $ violating parameter in the semileptonic amplitude,
$ \Rp{y_l} $, will be determined to $ \pm 0.19 \times 10^{-3} $.
 Inclusive semileptonic decay channel also gives
$ \Ip\Delta $ to $ \pm 0.18 \times 10^{-3} $ \cite{Buchanan}.
 The $ \dt = ( 0 \sim 15 ) \taus $ behavior of
$ {\cal A}^{l^+ l^-}(\dt) $
( see Eq.$ \left( \ref{ eqn:tll } \right) $ ) gives $ \Ip\BD $.
 If $ \Ip\BD $ can be measured
with the precision $ \pm 2.0 \times 10^{-3} $,
the precision of $ \Ip(x_l+\bar{x}_l) $ can reach the precision level
$ \pm 2.0 \times 10^{-3} $.
%
%
%
%
%
%
%
%%%%%%%%%%%%%%%%%%%%%%%%%%%%%%%%%%%%%%%%%%%%%%%%%%%%%%%%%%%%%%%%%
%                                                               %
%                         Appendix A                            %
%                                                               %
%%%%%%%%%%%%%%%%%%%%%%%%%%%%%%%%%%%%%%%%%%%%%%%%%%%%%%%%%%%%%%%%%
%
%
\vfill
\newpage
%
%{ \Large \bf Appendix }
%\section{ $ \Gamma_{ij} $ and $ M_{ij} $ }
%
%\renewcommand{\theequation}{ \Alph{section}.\arabic{equation} }
%\ap
%
\appendix{ }
 This appendix is devoted to giving the expressions of
the effective Hamiltonian
which describes $ K $ mesons' time evolution.
 They are used in checking the transformation propertities
of various parameters in the neutral $ K $ meson system
under discrete symmetries.
 Usual perturbative calculation in the second order with respect to
the small perturbative part $ H_W $ yields \cite{LeeWu}
\begin{eqnarray}
 M_{11} & = & m_K + \left< K^0 |H_W| K^0 \right> +
\displaystyle{
 \sum_n {\bf P}
 \frac{ \left| \left< n |H_W| K^0 \right> \right|^2 }
      { m_K - m_n }
}, \nonumber \\
 M_{22} & = & m_K + \left< \bar{K}^0 |H_W| \bar{K}^0 \right> +
\displaystyle{
 \sum_n {\bf P}
 \frac{ \left| \left< n |H_W| \bar{K}^0 \right> \right|^2 }
      { m_K - m_n }
}, \\
 M_{12} & = & \left< K^0 | H_W | \bar{K}^0 \right> +
\displaystyle{
 \sum_n {\bf P}
 \frac{ \left< K^0 |H_W| n \right> \left< n |H_W| \bar{K}^0 \right> }
      { m_K - m_n }
},\nonumber
\end{eqnarray}
 where {\bf P} stands for taking the principal part, and $ m_n $ is
the energy of the state $ \left| n \right> $
in the $ K $ meson rest frame.
 As for $ \Gamma_{ij}^\prime $s,
\begin{eqnarray}
 \Gamma_{11} & = & 2 \pi \sum_n \delta ( m_K - m_n )
\left| \left< n |H_W| K^0 \right> \right|^2, \nonumber \\
 \Gamma_{22} & = & 2 \pi \sum_n \delta ( m_K - m_n )
\left| \left< n |H_W| \bar{K}^0 \right> \right|^2, \\
 \Gamma_{12} & = & 2 \pi \sum_n \delta ( m_K - m_n )
\left< K^0 |H_W| n \right> \left< n |H_W| \bar{K}^0 \right>
\nonumber.
\end{eqnarray}
 With the use of the above expressions,
we can verify in the physical phase convention that
\begin{itemize}
 \item $ CPT $ conservation $ \Longrightarrow $ $ M_{11} = M_{22} $
  and $ \Gamma_{11} = \Gamma_{22} $
 \item $ CP $ conservation $ \Longrightarrow $ $ M_{11} = M_{22},
  \Gamma_{11} = \Gamma_{22},\ \Ip M_{12} = 0 $ and $ \Ip \Gamma_{12}
   = 0$
 \item $ T $ conservation $ \Longrightarrow $ $ \Ip M_{12} = 0 $
  and $ \Ip \Gamma_{12} = 0 $.
\label{ eqn:Width }
\end{itemize}
%
%
%%%%%%%%%%%%%%%%%%%%%%%%%%%%%%%%%%%%%%%%%%%%%%%%%%%%%%%%%%%%%%%%%%%
%                                                                 %
%                            Appendix B                           %
%                                                                 %
%%%%%%%%%%%%%%%%%%%%%%%%%%%%%%%%%%%%%%%%%%%%%%%%%%%%%%%%%%%%%%%%%%%
%
%
\appendix{ }
 In this appendix the expressions of $ \epsilon $
and $ \epsilon^\prime $
in the Eq.$ \left( \ref{ eqn:epsilonprime } \right) $ are derived
with a few remarks.
 From Eqs.$ \left( \ref{ eqn:msrl }  \right), \left( \ref{ eqn:A }
  \right) $
we have
\begin{eqnarray}
 \left< (2\pi)_I |T| K_L \right> & = &
\displaystyle{
 \frac{1}{ \sqrt{ 2 ( 1 + \left| \epsilon_2 \right|^2 ) } }
}
e^{ i \delta_I } ( p_2 A_I - q_2 \bar{A}_I ), \nonumber \\
 \left< (2\pi)_I |T| K_S \right> & = &
\displaystyle{
 \frac{1}{ \sqrt{ 2 ( 1 + \left| \epsilon_1 \right|^2 ) } }
}
e^{ i \delta_I } ( p_1 A_I + q_1 \bar{A}_I ). \nonumber
\end{eqnarray}
 Using the relations between the charge eigenstates
$ \left| \pi^+ \pi^- \right>, \left| \pi^0 \pi^0 \right> $ and
isospin eigenstates $ \left| ( 2 \pi )_I \right> $
\begin{equation}
\begin{array}{rcl}
 \left| \pi^+ \pi^- \right> & = &
\displaystyle{
 \sqrt{ \frac{1}{3} } \left| ( 2 \pi )_2 \right>
+ \sqrt{ \frac{2}{3} } \left| ( 2 \pi )_0 \right>
}, \\
 \left| \pi^0 \pi^0 \right> & = &
\displaystyle{
 \sqrt{ \frac{2}{3} } \left| ( 2 \pi )_2 \right>
- \sqrt{ \frac{1}{3} } \left| ( 2 \pi )_0 \right>
},
\end{array}
\end{equation}
one get
\begin{equation}
\begin{array}{rcl}
 \left< \pi^+ \pi^- |T| K_L \right> & = &
\displaystyle{
 \frac{1}{ \sqrt{3} } e^{ i \delta_0 } p_2 A_0 \left\{
\left( 1 - \frac{ q_2 }{ p_2 } \frac{ \bar{A}_0 }{ A_0 } \right)
+ \frac{1}{ \sqrt{2} } e^{ i ( \delta_2 -  \delta_0 ) }
\frac{ p_2 A_2 - q_2 \bar{A}_2 }{ p_2 A_0 } \right\}
}, \\
 & & \\
 \left< \pi^0 \pi^0 |T| K_L \right> & = &
\displaystyle{
 - \frac{1}{ \sqrt{6} } e^{ i \delta_0 } p_2 A_0 \left\{
\left( 1 - \frac{ q_2 }{ p_2 } \frac{ \bar{A}_0 }{ A_0 } \right)
- \sqrt{2} e^{ i ( \delta_2 - \delta_0 ) }
\frac{ p_2 A_2 - q_2 \bar{A}_2 }{ p_2 A_0 } \right\}
},
\end{array}
\end{equation}
and by ignoring $ CP $ violation in $ K_S $ decay( it is sufficient
since the denominators in $ \left| \eta_{+-} \right| $ and
$ \left| \eta_{00} \right| $
are already order of $ CP, T $ or $ CPT $ violation
parameters ),
\begin{equation}
\begin{array}{rcl}
 \left< \pi^+ \pi^- |T| K_S \right> & = &
\displaystyle{
 \frac{2}{ \sqrt{3} } e^{ i \delta_0 } p_1 A_0 \left(
 1 + \frac{1}{ \sqrt{2} }
       \omega e^{ i( \delta_2 - \delta_0 ) } \right)
}, \\
 & & \\
 \left< \pi^0 \pi^0 |T| K_S \right> & = &
\displaystyle{
 - \frac{2}{ \sqrt{6} } e^{ i \delta_0 } p_1 A_0 \left(
 1 - \sqrt{2} \omega e^{ i ( \delta_2 - \delta_0 ) } \right)
}.
\end{array}
\end{equation}
 Numerically $ \omega \simeq \frac{1}{20} $.
Thus neglecting terms of order $ \omega^2 $, we have
\begin{equation}
\begin{array}{rcl}
 \eta_{+-} & = &
\displaystyle{
 \frac{1}{2}
   \left( 1 - \frac{ q_2 }{ p_2 } \frac{ \bar{A}_0 }{ A_0 } \right)
- \frac{1}{ 2 \sqrt{2} } \omega e^{ i ( \delta_2 - \delta_0 ) }
\left( 1 - \frac{ q_2 }{ p_2 } \frac{ \bar{A}_0 }{ A_0 } \right)
} \\
 & &
\displaystyle{
 + \frac{1}{ 2 \sqrt{2} } e^{ i ( \delta_2 - \delta_0 ) }
 \left( \frac{ p_2 A_2 - q_2 \bar{A}_2 }{ p_2 A_0 } \right)
}, \\
 & & \\
 \eta_{00} & = &
\displaystyle{
 \frac{1}{2}
   \left( 1 - \frac{ q_2 }{ p_2 } \frac{ \bar{A}_0 }{ A_0 } \right)
+ \frac{1}{ \sqrt{2} } \omega e^{ i ( \delta_2 - \delta_0 ) }
\left( 1 - \frac{ q_2 }{ p_2 } \frac{ \bar{A}_0 }{ A_0 } \right)
} \\
 & &
\displaystyle{
 - \frac{1}{ \sqrt{2} } e^{ i ( \delta_2 - \delta_0 ) }
 \left( \frac{ p_2 A_2 - q_2 \bar{A}_2 }{ p_2 A_0 } \right)
}.
\end{array}
\end{equation}
 Expressing the above equations
in terms of $ \epsilon $ and $ \epsilon^\prime $,
one gets the results in
Eq.$ \left( \ref{ eqn:epsilonepsilonprime } \right) $
\begin{equation}
\begin{array}{rcl}
 \epsilon & = &
\displaystyle{
\frac{1}{2} \left( 1
- \frac{ q_2 }{ p_2 } \frac{ \bar{A}_0 }{ A_0 } \right)
}, \\
 \epsilon^\prime & = &
\displaystyle{
 \frac{1}{ 2 \sqrt{2} } e^{ i ( \delta_2 - \delta_0 ) }
 \left[
  - \omega \left( 1 - \frac{ q_2 }{ p_2 } \right)
  + \left(
     \frac{ p_2 A_2 - q_2 \bar{A}_2 }{ p_2 A_0 }
    \right)
 \right]
}.
\end{array}
\label{ eqn:appepsilon }
\end{equation}
\par
 Note that we have not assumed $ CPT $ symmetry
in deriving the expressions
for $ \epsilon $ and $ \epsilon^\prime $.
 $ \epsilon $ measured in the $ K_L $ decay is sensitive
only to $ \epsilon_2 $.
 This indicates the importance of measuring $ CP $ violating effects
in the $ K_S $ decay.
 As noted in the Sec.\ref{ subsec:Phase },
the relative phase between $ CP \KO $ and $ \KB $
remains still unfixed.
 The Wu-Yang phase convention \cite{WuYan} is
to fix the phase of $ \KO $ such that
\begin{equation}
 \Ip A_0 = 0,
\end{equation}
with $ CP \KO = \KB $ and $ CPT \KO = \KB $.
 This convention is convenient only if $ CPT $ invariance is assumed,
since then $ \epsilon $ is exactly equal to $ \epsilon_0 $
\cite{LeeWu};
\begin{equation}
\begin{array}{rcl}
 \epsilon & = & \epsilon_0, \\
 \epsilon^\prime & = &
\displaystyle{
 \frac{i}{ \sqrt{2} } e^{ i ( \delta_2 - \delta_0 ) }
 \omega \frac{ \Ip A_2 }{ \Rp A_2 }
}.
\label{ eqn:WuYan }
\end{array}
\end{equation}
 This is the standard phase convention used by many authors
in the context of $ CP $ violation.
 We find, however, it is more convenient to use the expressions
defined in Eq.$ \left( \rm{\ref{ eqn:appepsilon }} \right) $,
as there is no simplification by taking Wu-Yang convention.
%
%
%%%%%%%%%%%%%%%%%%%%%%%%%%%%%%%%%%%%%%%%%%%%%%%%%%%%%%%%%%%%%%%
%                                                             %
%                          Appendix C                         %
%                                                             %
%%%%%%%%%%%%%%%%%%%%%%%%%%%%%%%%%%%%%%%%%%%%%%%%%%%%%%%%%%%%%%%
%
%
\appendix{ }
\label{ app:prob }
 This appendix contains the expressions for
$ \left| A( f_1, t_1; f_2, t_2 ) \right|^2 $ for
$ f_1,\, f_2 = \pi^- l^+ \nu_l $,
$ \pi^+ l^- \bar{\nu}_l, \pi^+ \pi^- $ or
$ \pi^0 \pi^0 $.
 They were obtained in Ref.\cite{Buchanan}
in the case that $ x_l, \bar{x}_l = 0 $.
 Each probability is the sum of three time dependence;
$ e^{ -( \wl t_1 + \ws t_2 ) },\ e^{ -( \ws t_1 + \wl t_2 ) }, $
and $ e^{ -\frac{\wt}{2} (t_1 + t_2) }
\times\{ \sin \ {\rm or}\  \cos \} $.
 Coefficients were calculated to the first order with respect to
small parameters like $ \epsilon_0 $.
 Exceptional case is the one in which such an approximation gives
0 for the value of some coefficient.
 Then the second order contribution to it was calculated.
\begin{eqnarray}
 \left| A( \pi \pi, t_1; \pi \pi, t_2 ) \right|^2 & = &
\displaystyle{
 \frac{1}{2} \left| \left< \pi \pi |T| K_S \right> \right|^2
\left| \left< \pi \pi |T| K_L \right> \right|^2
} \nonumber \\
 & & \times \Bigl[ e^{ -( \wl t_1 + \ws t_2 ) }
+ e^{ -( \ws t_1 + \wl t_2 ) }
\nonumber \\
 & & \quad \quad \quad \quad - 2 e^{ -\frac{\wt}{2} ( t_1 + t_2 ) }
\cos \left( \dm( t_1 - t_2 ) \right) \Bigl],
\end{eqnarray}
\begin{eqnarray}
 \lefteqn{ \left| A( \pi^+ \pi^-, t_1; \pi^0 \pi^0, t_2 ) \right|^2 }
\nonumber \\
 & = &
\displaystyle{
 \frac{1}{2} \left| \left< \pi^+ \pi^- |T| K_S \right> \right|^2
\left| \left< \pi^0 \pi^0 |T| K_S \right> \right|^2
} \nonumber \\
 & & \times \Bigl[ \left| \eta_{\pi^+ \pi^-} \right|^2
e^{ -( \wl t_1 + \ws t_2 ) }
+ \left| \eta_{\pi^0 \pi^0} \right|^2 e^{ -( \ws t_1 + \wl t_2 ) }
\nonumber \\
 & & - 2 \left| \eta_{\pi^+ \pi^-} \right| \left| \eta_{\pi^0 \pi^0}
         \right|
e^{ -\frac{\wt}{2} ( t_1 + t_2 ) } \cos \left( \dm( t_1 - t_2 ) -
   \phi_{\pi^+ \pi^-}
+ \phi_{\pi^0 \pi^0} \right) \Bigl], \nonumber \\
\end{eqnarray}
\begin{eqnarray}
 \lefteqn{
    \left| A( \pi^- l^+ \nu_l, t_1 ; \pi^+ l^- \bar{\nu}_l, t_2 )
    \right|^2 }
\nonumber \\
 & = &
\displaystyle{
 2 \left| \frac{ F_l }{2} \right|^4 \Bigl[
( 1 + 4 \Rp \BD ) e^{ - ( \ws t_1 + \wl t_2 ) }
}
 + ( 1 - 4 \Rp \BD ) e^{ -( \wl t_1 + \ws t_2 ) } \nonumber \\
& & + 2 e^{ -\frac{\wt}{2} ( t_1 + t_2 ) }
              \left\{ \cos( \dm ( t_1 - t_2 ) )
                - 4 \Ip \BD \sin( \dm ( t_1 - t_2 ) )
              \right\} \Bigl],
\end{eqnarray}
\begin{eqnarray}
 \lefteqn{
  \left|
   A( \pi^+ l^- \bar{\nu}_l, t_1 ; \pi^+ l^- \bar{\nu}_l, t_2 )
  \right|^2 }
  \nonumber \\
 & = &
\displaystyle{
 2 \left| \frac{ F_l }{2} \right|^4
\left( 1 - 4 \Rp( \epsilon_0 - y ) \right)
\Bigl[
e^{ - ( \ws t_1 + \wl t_2 ) } +  e^{ -( \wl t_1 + \ws t_2 ) }
} \nonumber \\
 & & - 2 e^{ -\frac{\wt}{2} ( t_1 + t_2 ) } \cos( \dm ( t_1 - t_2 ) )
\Bigl],
\end{eqnarray}
\begin{eqnarray}
 \lefteqn{ \left| A( \pi^- l^+ \nu_l, t_1 ; \pi^- l^+ \nu_l, t_2 )
\right|^2 }
\nonumber \\
 & = &
\displaystyle{
 2 \left| \frac{ F_l }{2} \right|^4
\left( 1 + 4 \Rp( \epsilon_0 - y ) \right)
\Bigl[
e^{ - ( \ws t_1 + \wl t_2 ) } +  e^{ -( \wl t_1 + \ws t_2 ) }
} \nonumber \\
 & & - 2 e^{ -\frac{\wt}{2} ( t_1 + t_2 ) } \cos( \dm ( t_1 - t_2 ) )
\Bigl],
\end{eqnarray}
\begin{eqnarray}
 \lefteqn{ \left| A( \pi^- l^+ \nu_l, t_1; \pi\pi, t_2 ) \right|^2 }
\nonumber \\
 & = &
\displaystyle{
 \left| \frac{ F_l }{2} \right|^2
\left| \left< \pi\pi |T| K_S \right> \right|^2 \Bigl[
\left| \eta_{\pi\pi} \right|^2 e^{ -( \ws t_1 + \wl t_2 ) }
} \nonumber \\
 & + & \left( 1 + 2 \Rp( \epsilon_0 - y_l )
       - 2 \Rp( \Delta + x_l ) \right)
         e^{ -( \wl t_1 + \ws t_2 ) } \nonumber \\
 & - & 2 \left| \eta_{\pi\pi} \right|
         e^{ -\frac{\wt}{2} ( t_1 + t_2 ) }
\cos \left( \dm (t_1- t_2) + \phi_{\pi\pi} \right) \Bigl],
\end{eqnarray}
\begin{eqnarray}
 \lefteqn{ \left| A( \pi^+ l^- \bar{\nu}_l, t_1; \pi\pi, t_2 )
           \right|^2 }
\nonumber \\
 & = &
\displaystyle{
 \left| \frac{ F_l }{2} \right|^2
\left| \left< \pi\pi |T| K_S \right> \right|^2 \Bigl[
\left| \eta_{\pi\pi} \right|^2 e^{ -( \ws t_1 + \wl t_2 ) }
} \nonumber \\
 & + & \left( 1 - 2 \Rp( \epsilon_0 - y_l )
    + 2 \Rp( \Delta - \bar{x}_l )
\right) e^{ -( \wl t_1 + \ws t_2 ) } \nonumber \\
 & + & 2 \left| \eta_{\pi\pi} \right|
      e^{ -\frac{\wt}{2} ( t_1 + t_2 ) }
\cos \left( \dm (t_1- t_2) + \phi_{\pi\pi} \right) \Bigl],
\end{eqnarray}
%
%
%%%%%%%%%%%%%%%%%%%%%%%%%%%%%%%%%%%%%%%%%%%%%%%%%%%%%%%%%%%%%
%                                                           %
%            Inclusive \pi\pi time distribution             %
%                                                           %
%%%%%%%%%%%%%%%%%%%%%%%%%%%%%%%%%%%%%%%%%%%%%%%%%%%%%%%%%%%%%
%
%
\begin{eqnarray}
 \left| A( \pi\pi, t ) \right|^2 & \equiv &
\sum_f \int_0^\infty dt^\prime \left| A( \pi\pi,t; f,t^\prime )
\right|^2            \nonumber \\
 & = &
\displaystyle{
 \frac{1}{2} \left| \left< \pi\pi |T| K_S \right> \right|^2 \Bigl\{
\left| \eta_{\pi\pi} \right|^2 e^{ -\wl t } + e^{ -\ws t }
}
\nonumber \\
 & & -
2 \left| \eta_{\pi\pi} \right| \left| \left< K_S | K_L \right>
  \right| e^{ -\frac{\wt}{2} t }
  \cos( \dm t - \phi_{\pi\pi} - \phi_{LS} ) \Bigl\},
\end{eqnarray}
%
%%%%%%%%%%%%%%%%%%%%%%%%%%%%%%%%%%%%%%%%%%%%%%%%%%%%%%%%%%%%%%
%                                                            %
%             Inclusive semileptonic decay mode              %
%                                                            %
%%%%%%%%%%%%%%%%%%%%%%%%%%%%%%%%%%%%%%%%%%%%%%%%%%%%%%%%%%%%%%
%
\begin{eqnarray}
 \left| A( \pi^- l^+ \nu_l,t ) \right|^2 & \equiv &
\sum_f \int_0^\infty dt^\prime
\left| A( \pi^- l^+ \nu, t; f, t^\prime ) \right|^2
\nonumber \\
 & = &
\displaystyle{
 \left| \frac{ F_l }{2} \right|^2 \Bigl[
\{ 1 + 2[ \Rp( \epsilon_0 - y_l ) - \Rp( \Delta + x_l ) ] \}
e^{ -\wl t }
}
\nonumber \\
 & & + \{ 1 + 2[ \Rp( \epsilon_0 - y_l ) + \Rp( \Delta + x_l ) ] \}
e^{ -\ws t } \nonumber \\
 & & - 2 \left| \left< K_S | K_L \right> \right|
e^{ -\frac{\wt}{2} t } \cos( \dm t - \phi_{LS} ) \Bigl],
\label{ eqn:inantilepto }
\end{eqnarray}
\begin{eqnarray}
 \left| A( \pi^+ l^- \bar{\nu}_l,t ) \right|^2 & \equiv &
\sum_f \int_0^\infty dt^\prime \left| A( \pi^+ l^- \bar{\nu}, t;
f, t^\prime )
\right|^2
\nonumber \\
 & = &
\left| \frac{ F_l }{2} \right|^2 \Bigl[
\{ 1 - 2[ \Rp( \epsilon_0 - y_l ) - \Rp( \Delta - \bar{ x_l } ) ] \}
e^{ -\wl t }
\nonumber \\
 & & + \{ 1 - 2[ \Rp( \epsilon_0 - y_l )
     + \Rp( \Delta - \bar{ x_l } ) ] \} e^{ -\ws t }
\nonumber \\
 & & + 2 \left| \left< K_S | K_L \right> \right|
e^{ -\frac{\wt}{2} t } \cos( \dm t - \phi_{LS} ) \Bigl].
\label{ eqn:inlepto }
\end{eqnarray}
 They may become useful
when more practical experimental situation is considered,
for example, the fiducial volume effect,
finiteness in the vertex resolution ability,
or experimental cut for small $ \left| \dt \right| $ value,
or when the experiments do not concentrate on the use of
$ \left| A( \dt; f_1, f_2 ) \right|^2 $.
%
%
%%%%%%%%%%%%%%%%%%%%%%%%%%%%%%%%%%%%%%%%%%%%%%%%%%%%%%%%%%%%%%%%%
%                                                               %
%                          Appendix D                           %
%                                                               %
%%%%%%%%%%%%%%%%%%%%%%%%%%%%%%%%%%%%%%%%%%%%%%%%%%%%%%%%%%%%%%%%%
%
%
\appendix{ }
\label{ app:relprob }
 In this appendix we summarize various relative time probability
distribution functions which appear at a $ \phi $ factory.
 The results are the same in Ref.\cite{Buchanan}
for the purely hadronic decay mode:
\begin{eqnarray}
 \lefteqn{ \left| A( \dt; \pi \pi, \pi \pi ) \right|^2 }
\nonumber \\
 & = & \displaystyle{
      \frac{1}{2 \wt}
     }
     \left| \left< \pi \pi |T| K_S \right>
     \left< \pi \pi |T| K_L \right> \right|^2
     \{ e^{- \wl \adt } + e^{ -\ws \adt }
\nonumber \\
 & & - 2 e^{ -\frac{\wt}{2} \adt } \cos ( \dm \adt ) \},
\label{ eqn:samepipi }
\end{eqnarray}
\begin{eqnarray}
 \lefteqn{ \left| A( \dt>0; \pi^+ \pi^-, \pi^0 \pi^0 ) \right|^2 }
\nonumber \\
 & = & \displaystyle{
          \frac{ \left| \epsilon \right|^2 }{2 \wt}
         }
     \left | \left< \pi^+ \pi^- |T| K_S \right>
     \left< \pi^0 \pi^0 |T| K_S \right> \right|^2
\nonumber \\
 & & \Bigl[
 ( 1 + 2 \Rp ( \epsilon^\prime /\epsilon ) ) e^{ -\wl \dt }
+ ( 1 - 4 \Rp ( \epsilon^\prime /\epsilon ) ) e^{ -\ws \dt }
\nonumber \\
 & & - 2 e^{ -\frac{\wt}{2} \dt }
\{ ( 1 - \Rp ( \epsilon^\prime /\epsilon ) ) \cos ( \dm \dt )
+ 3 \Ip ( \epsilon^\prime /\epsilon ) \sin ( \dm \dt ) \} \Bigl],
\end{eqnarray}
\begin{eqnarray}
 \lefteqn{ \left| A( \dt<0; \pi^+ \pi^-, \pi^0 \pi^0 ) \right|^2 }
\nonumber \\
 & = & \displaystyle{
          \frac{ \left| \epsilon \right|^2 }{2 \wt}
         }
     \left| \left< \pi^+ \pi^- |T| K_S \right>
     \left< \pi^0 \pi^0 |T| K_S \right> \right|^2
\nonumber \\
 & & \Bigl[
 ( 1 - 4 \Rp ( \epsilon^\prime /\epsilon ) ) e^{ -\wl\adt }
+ ( 1 + 2 \Rp ( \epsilon^\prime /\epsilon ) ) e^{ -\ws\adt }
\nonumber \\
 & & - 2 e^{ -\frac{\wt}{2} \adt }
\{ ( 1 - \Rp ( \epsilon^\prime /\epsilon ) ) \cos ( \dm \adt )
- 3 \Ip ( \epsilon^\prime /\epsilon ) \sin ( \dm \adt ) \} \Bigl].
\end{eqnarray}
 When at least one of two channels is semileptonic one, we have
\begin{eqnarray}
 \lefteqn{ \left| A( \dt>0 ; \pi^- l^+ \nu_l, \pi^+ l^- \bar{\nu}_l )
\right|^2 } \nonumber \\
 & = & \displaystyle{
        \frac{2}{\wt} \left| \frac{F_l}{2} \right|^4
       }
\Bigl[
( 1 - 4 \Rp ( \BD ) ) e^{ -\wl\dt } + ( 1 + 4 \Rp( \BD ) )
  e^{-\ws\dt }
\nonumber \\
 & & + 2 e^{ -\frac{\wt}{2} \dt } \{ \cos( \dm\dt ) - 4 \Ip ( \BD )
\sin( \dm\dt ) \} \Bigl],
\end{eqnarray}
\begin{eqnarray}
 \lefteqn{ \left| A( \dt<0 ; \pi^- l^+ \nu_l, \pi^+ l^- \bar{\nu}_l )
\right|^2 } \nonumber \\
 & = & \displaystyle{
        \frac{2}{\wt} \left| \frac{F_l}{2} \right|^4
       }
\Bigl[
( 1 + 4 \Rp ( \BD ) ) e^{ -\wl\adt } + ( 1 - 4 \Rp( \BD ) )
e^{-\ws\adt }   \nonumber \\
 & & + 2 e^{ -\frac{\wt}{2} \adt }
     \{ \cos( \dm\adt ) + 4 \Ip ( \BD ) \sin( \dm\adt ) \} \Bigl],
\end{eqnarray}
\begin{eqnarray}
 \lefteqn{
   \left|
       A( \dt; \pi^+ l^- \bar{\nu}_l, \pi^+ l^- \bar{\nu}_l )
   \right|^2 } \nonumber \\
 & = &
\displaystyle{
 \frac{2}{\wt} \left| \frac{ F_l }{2} \right|^4
}
( 1 - 4 \Rp ( \epsilon_0 - y_l ) )
\Bigl\{ e^{ -\wl\adt } + e^{ -\ws\adt }
\nonumber \\
 & & - 2 e^{ -\frac{\wt}{2} \adt } \cos( \dm\adt ) \Bigl\},
\end{eqnarray}
\begin{eqnarray}
 \lefteqn{ \left| A( \dt; \pi^- l^+ \nu_l, \pi^- l^+ \nu_l )
           \right|^2 }
\nonumber \\
 & = &
\displaystyle{
 \frac{2}{\wt} \left| \frac{ F_l }{2} \right|^4
}
( 1 + 4 \Rp ( \epsilon_0 - y_l ) )
\Bigl\{ e^{ -\wl\adt } + e^{ -\ws\adt }
\nonumber \\
 & & - 2 e^{ -\frac{\wt}{2} \adt } \cos( \dm\adt ) \Bigl\},
\end{eqnarray}
\begin{eqnarray}
 \lefteqn{ \left| A( \dt>0; \pi^- l^+ \nu_l, \pi\pi ) \right|^2 }
\nonumber \\
 & = &
\displaystyle{
 \frac{1}{\wt} \left| \frac{F_l}{2} \right|^2
}
\left| \left< \pi\pi |T| K_S \right> \right|^2
\nonumber \\
 & & \times
\Bigl[
\Bigl\{ 1 + 2 \left[ \Rp( \epsilon_0 - y_l ) - \Rp( \Delta + x_l )
              \right]
\Bigl\} e^{ -\wl\dt }
 + \left| \eta_{\pi\pi} \right|^2 e^{ -\ws\dt }
\nonumber \\
 & & - 2 \left| \eta_{\pi\pi} \right| e^{ -\frac{\wt}{2} \dt }
\cos( \dm\dt + \phi_{\pi\pi} )
\Bigl],
\end{eqnarray}
\begin{eqnarray}
 \lefteqn{ \left| A( \dt<0; \pi^- l^+ \nu_l, \pi\pi ) \right|^2 }
\nonumber \\
 & = &
\displaystyle{
 \frac{1}{\wt} \left| \frac{ F_l }{2} \right|^2
\left| \left< \pi\pi |T| K_S \right> \right|^2
}
\nonumber \\
 & & \times
\Bigl[
\left| \eta_{\pi\pi} \right|^2 e^{ -\wl\adt }
+ \Bigl\{ 1 + 2 \left[ \Rp( \epsilon_0 - y_l ) - \Rp( \Delta + x_l )
                \right] \Bigl\} e^{ -\ws\adt }
\nonumber \\
 & & - 2 \left| \eta_{\pi\pi} \right| e^{ -\frac{\wt}{2} \adt }
\cos( \dm\adt - \phi_{\pi\pi} )
\Bigl],
\end{eqnarray}
\begin{eqnarray}
 \lefteqn{ \left| A( \dt>0; \pi^+ l^- \bar{\nu}_l, \pi\pi )
           \right|^2 }
\nonumber \\
 & = &
\displaystyle{
 \frac{1}{\wt} \left| \frac{ F_l }{2} \right|^2
\left| \left< \pi\pi |T| K_S \right> \right|^2
}
\nonumber \\
 & & \times
\Bigl[
 \Bigl\{ 1 - 2 \left[ \Rp( \epsilon_0 - y_l )
        - \Rp( \Delta - \bar{x_l} )
 \right] \Bigl\} e^{ -\wl\dt }
 + \left| \eta_{\pi\pi} \right|^2 e^{ -\ws\dt }
\nonumber \\
 & & + 2 \left| \eta_{\pi\pi} \right| e^{ -\frac{\wt}{2} \dt}
\cos( \dm\dt + \phi_{\pi\pi} )
\Bigl],
\end{eqnarray}
\begin{eqnarray}
 \lefteqn{ \left| A( \dt<0; \pi^+ l^- \bar{\nu}_l, \pi\pi )
           \right|^2 }
\nonumber \\
 & = &
\displaystyle{
 \frac{1}{\wt} \left| \frac{ F_l }{2} \right|^2
}
\left| \left< \pi\pi |T| K_S \right> \right|^2
\nonumber \\
 & & \times
\Bigl[
\left| \eta_{\pi\pi} \right|^2 e^{ -\wl\adt }
+ \Bigl\{ 1 - 2 \left[ \Rp( \epsilon_0 - y_l )
       - \Rp( \Delta - \bar{x_l} )
  \right] \Bigl\} e^{ -\ws\adt }
\nonumber \\
 & & + 2 \left| \eta_{\pi\pi} \right| e^{ -\frac{\wt}{2} \adt }
\cos( \dm\adt - \phi_{\pi\pi} )
\Bigl].
\end{eqnarray}
\vfill
\newpage
%
%
%%%%%%%%%%%%%%%%%%%%%%%%%%%%%%%%%%%%%%%%%%%%%%%%%%%%%%%%%%%%%%%%%%%
%                                                                 %
%                              Reference                          %
%                                                                 %
%%%%%%%%%%%%%%%%%%%%%%%%%%%%%%%%%%%%%%%%%%%%%%%%%%%%%%%%%%%%%%%%%%%
%
%

%
%
%
%
%
%%%%%%%%%%%%%%%%%%%%%%%%%%%%%%%%%%%%%%%%%%%%%%%%%%%%%%%%%%%%%%
%                                                            %
%                         Table 1                            %
%                                                            %
%%%%%%%%%%%%%%%%%%%%%%%%%%%%%%%%%%%%%%%%%%%%%%%%%%%%%%%%%%%%%%
%
%
\widetext
\begin{table}
\caption{ The measurement of parameters at a $ \phi $ factory.
          The Eq.number in the right of each quantity shows
         that appearing in the text.
          The $ \dt $ or $ t $ region indicates that of the quantity
         which is sensitive to the parameter we want.
          $ \delta_l(\infty) $ and $ \bar{\delta}_l(\infty) $
         are $ \delta_l(\infty) = 2( \Rp( \epsilon_0 - y_l )
         - \Rp \BD ) $
         and $ \bar{\delta}_l(\infty) = 2( \Rp( \epsilon_0 - y_l )
         + \Rp \BD )$ respectively.
          We complete this table
         by the addition of the results obtained
         in Ref.\cite{Buchanan} }
\begin{tabular}{cccl}
 parameter & quantity & Eq.number & $ \dt $ or $ t $ region
\\ \tableline
 $ \ws $ & $ \left| A( \pi^+ \pi^-, t) \right|^2 $
& Eq.$ \left( \ref{ eqn:inpipi } \right) $ &
   $ t = ( 0 \sim 15 ) \taus $
\\ \tableline
 $ \wl $
&
 \begin{tabular}{c}
  $ A( \pi^- l^+ \nu_l, t ) $ \\
  $ A( \pi^+ l^- \bar{\nu}_l, t ) $
 \end{tabular}
&
 \begin{tabular}{c}
  Eq.$ \left( \rm{ \ref{ eqn:inantilepto } } \right) $ \\
  Eq.$ \left( \rm{ \ref{ eqn:inlepto } } \right) $ \\
 \end{tabular}
&
 \begin{tabular}{c}
  $ t \gg  \taus $ \\
  $ t \gg  \taus $
 \end{tabular}
\\ \tableline
 $ \dm $ & $ A( \dt; \pi^+ \pi^-, \pi^+ \pi^- ) $
& Eq.$ \left( \rm{ \ref{ eqn:samepipi } } \right) $
& $ \dt = ( 0 \sim 10 ) \taus $
\\ \tableline
 $ \phi_{SW} $ & From $ {\rm Arctan}( 2 \dm/(\ws-\wl) ) $ & &
\\ \tableline
 $ \left| \eta_{+-} \right| $ & $ \left| A(\pi^+ \pi^-,t) \right|^2 $
& Eq.$ \left( \ref{ eqn:inpipi } \right) $
& $ t \gg  \taus $
\\ \tableline
 $ \left| \eta_{00} \right| $ & $ \left| A(\pi^0 \pi^0,t) \right|^2 $
& Eq.$ \left( \ref{ eqn:inpipi } \right) $
& $ t \gg  \taus $
\\ \tableline
 $ \phi_{+-} $
&
 \begin{tabular}{c}
  $ {\cal A}^-_{\pi^+ \pi^-,l^+ l^-} $ \\
  $ {\cal A}^{\pi^+ \pi^-}
     ( \dt<0; \pi^- l^+ \nu_l, \pi^+ l^- \bar{\nu}_l ) $
 \end{tabular}
&
 \begin{tabular}{c}
  Eq.$ \left( \ref{ eqn:cmppll } \right) $ \\
  Eq.$ \left( \ref{ eqn:tppmll } \right) $
 \end{tabular}
&
 \begin{tabular}{c}
  \\
  $ \left| \dt \right| = ( 0 \sim 15 ) \taus $
 \end{tabular}
\\ \tableline
 $ \Rp( \epsilon^\prime /\epsilon ) $
&
 \begin{tabular}{c}
  $ {\cal A}^{\pi^+ \pi^-, \pi^0 \pi^0} $ \\
  $ {\cal A} ( \dt; \pi^+ \pi^-, \pi^0 \pi^0) $ \\
  double ratio $ R $
 \end{tabular}
&
 \begin{tabular}{c}
  Eq. $ \left( \ref{ eqn:pipi } \right) $ \\
  Eq. $ \left( \ref{ eqn:tpipi } \right) $ \\

 \end{tabular}
&
 \begin{tabular}{c}
   \\
  $ \dt \gg 20 \taus $ \\

 \end{tabular}
\\ \tableline
 $ \Ip( \epsilon^\prime /\epsilon ) $
& $ {\cal A} ( \dt; \pi^+ \pi^-, \pi^0 \pi^0 ) $
& Eq.$ \left( \ref{ eqn:tpipi } \right) $
& $ \dt = ( 0 \sim 15 ) \taus $
\\ \tableline
 $ ( \phi_{+-} - \phi_{00} ) $
& From $ 3 \Ip( \epsilon^\prime /\epsilon ) $ & &
\\ \tableline
 $ \epsilon $ & From $ \frac{1}{3} ( \eta_{+-} + 2 \eta_{00} ) $ & &
\\ \tableline
 ( Table \ref{ tab:phi } continued ) & & &
\\ \tableline
 $ \delta_l (\infty) $
&
 \begin{tabular}{c}
  $ {\cal A}^{\pi\pi}
    (\Delta t>0; \pi^- l^+ \nu_l, \pi^+ l^- \bar{\nu}_l ) $ \\
  $ {\cal A}^+_{\pi\pi, l^+l^-} $ \\
  $ {\cal A}^{\rm inclusive}_{l^+ l^-} (t) $ \\
  $ {\cal A}^{\rm inclusive}_{l^+ l^-} $
 \end{tabular}
&
 \begin{tabular}{c}
  $ {\rm Eq.} \left( \ref{ eqn:tpppll } \right) $ \\
  $ {\rm Eq.} \left( \ref{ eqn:cpppll } \right) $ \\
  $ {\rm Eq.} \left( \ref{ eqn:tinll }  \right) $ \\
  $ {\rm Eq.} \left( \ref{ eqn:cinll } \right) $
 \end{tabular}
&
 \begin{tabular}{l}
  $ \dt \gg \taus $ \\
  \\
  $ t \gg \taus $ \\

 \end{tabular}
\\ \tableline
 $  \bar{ \delta }_l (\infty) $
&
 \begin{tabular}{l}
  Measurement of $ K_S $ decay \\
  \quad with $ K_L $ tagging
 \end{tabular}
& &
\\ \tableline
 \( \Rp \BD \)
&
 \begin{tabular}{c}
  $ {\cal A}^{l^+ l^-} ( \Delta t ) $ \\
  $ {\cal A}_{l^+ l^-} $ \\
  $ {\rm From} \  - \frac{1}{4}
    \left( \delta_l (\infty) - \bar{\delta}_l (\infty) \right) $
 \end{tabular}
&
 \begin{tabular}{c}
  $ {\rm Eq.} \left( \ref{ eqn:tll } \right) $ \\
  $ {\rm Eq.} \left( \ref{ eqn:cll } \right) $ \\

 \end{tabular}
&
 \begin{tabular}{l}
  $ \dt \gg \taus $ \\
   \\

  \end{tabular}
\\ \tableline
 \( \Ip \BD \)
&
 $ {\cal A}^{l^+ l^-} ( \dt ) $
&
 $ {\rm Eq.} \left( \ref{ eqn:tll } \right) $
&
 $ \dt = ( 0 \sim 15 ) \taus $
\\ \tableline
 \( \Rp( \epsilon_0 - y_l ) \)
&
 \begin{tabular}{c}
  $ {\cal A}( \Delta t; \pi^+ l^- \bar{\nu}_l, \pi^- l^+ \nu_l ) $ \\
  $ {\cal A}( \pi^+ l^- \bar{\nu}_l, \pi^- l^+ \nu_l ) $ \\
  $ {\rm From} \  \frac{1}{4}
    \left( \delta_l (\infty) + \bar{\delta}_l (\infty) \right) $
 \end{tabular}
&
 \begin{tabular}{c}
  $ {\rm Eq.} \left( \ref{ eqn:tplpl } \right) $ \\
  $ {\rm Eq.} \left( \ref{ eqn:cplpl } \right) $ \\

 \end{tabular}
&
\\ \tableline
 \( \Rp \epsilon_0 \)
&
 $ {\cal A}^{\rm inclusive}_{l^+ l^-} (t) $
&
 $ {\rm Eq.} \left( \ref{ eqn:tinll } \right) $
&
 $ t=( 0 \sim 15 ) \taus $
\\ \tableline
 \( \Ip \Delta \)
&
 $ {\cal A}^{\rm inclusive}_{l^+ l^-} (t) $
&
 $ {\rm Eq.} \left( \ref{ eqn:tinll } \right)$
&
 $ t=( 0 \sim 15) \taus $
\\ \tableline
 $ \Rp y_l $
&
 From  $ \Rp \epsilon_0 - \Rp( \epsilon_0 - y_l ) $
& &
\\ \tableline
 $ \Ip( x_l + \bar{x}_l ) $ & $ {\rm From} \  \Ip \BD - \Ip \Delta $
& &
\\ \tableline
 $ \frac{ \Ip A_0 }{ \Rp A_0 }
                 - \frac{ \Ip \bar{A}_0 }{ \Rp A_0 } $
& From $ 2 \Ip( \epsilon - ( \epsilon_0 - \Delta ) ) $ & &
\\ \tableline
$ \frac{ \Ip A_2 }{ \Rp A_2 }
                 - \frac{ \Ip \bar{A}_2 }{ \Rp A_2 } $
&
 $
 \begin{array}{l}
  {\rm From} \
  \Ip\left(
      \frac{ 2 \sqrt{2} \epsilon^\prime }{ \omega }
       e^{ -i( \delta_2 - \delta_0 ) }
     \right) \\
  \quad \quad \quad \quad + \left(
                             \frac{ \Ip A_0 }{ \Rp A_0 }
                             - \frac{ \Ip \bar{A}_0 }{ \Rp A_0 }
                            \right)
 \end{array}
 $
& &
\end{tabular}
\label{ tab:phi }
\end{table}
%
%
%%%%%%%%%%%%%%%%%%%%%%%%%%%%%%%%%%%%%%%%%%%%%%%%%%%%%%%%%%%%%%%%%%%%
%                                                                  %
%                             Table 2                              %
%                                                                  %
%%%%%%%%%%%%%%%%%%%%%%%%%%%%%%%%%%%%%%%%%%%%%%%%%%%%%%%%%%%%%%%%%%%%
%
%
\begin{table}
\caption{
 The coefficients $ A-D $ in the time dependent asymmetries
$ \beta_l(t) $ ( Eq.$ \left( \ref{ eqn:asybeta } \right) $)
and $ \alpha_l(t) $ ( Eq.$ \left( \ref{ eqn:asyalpha } \right) $).
}
\begin{tabular}{ccccc}
 & $ A $ & $ B $ & $ C $ & $ D $ \\
\tableline
 \( \alpha_l \)
 &
$\begin{array}{l}
 2 \Rp y_l + 4 \Rp \Delta \\
 + \Rp( x_l - \bar{x}_l )
\end{array}$
&
$\begin{array}{l}
 2 \Rp y_l - 4 \Rp \Delta \\
 - \Rp( x_l - \bar{x}_l )
\end{array}$
& \( 4 \Rp y_l \)
&
$\begin{array}{l}
 8 \Ip \Delta \\
 + 2 \Ip( x_l + \bar{x}_l )
\end{array}$
\\ \tableline
 \( \beta_l \)
&
$\begin{array}{l}
 4 \Rp \epsilon_0 - 2 \Rp y_l \\
 - \Rp( x_l - \bar{x}_l )
\end{array}$
 &
$\begin{array}{l}
 4 \Rp \epsilon_0 - 2 \Rp y_l \\
 + \Rp( x_l - \bar{x}_l )
\end{array}$
 & \( -4 \left[ 2 \Rp \epsilon_0 - \Rp y_l \right] \)
 & \( 2 \Ip( x_l + \bar{x}_l ) \)
\\
\end{tabular}
\label{ tab:abcoeff }
\end{table}
%
%%%%%%%%%%%%%%%%%%%%%%%%%%%%%%%%%%%%%%%%%%%%%%%%%%%%%%%%%%%%%%
%                                                            %
%                          Table 3                           %
%                                                            %
%%%%%%%%%%%%%%%%%%%%%%%%%%%%%%%%%%%%%%%%%%%%%%%%%%%%%%%%%%%%%%
%
%
\begin{table}
\caption{ The determination of the values of parameters
         from $ \beta_l(t) $ and $ \alpha_l(t) $ in the experiments
         with kaon tagging.
          The symbol( $ \alpha_l(t) $, $ A $ ) in the second column,
         for example, shows
         that the parameter can be determined from the measurement
         of the coefficient $ A $ in the asymmetry $ \alpha_l $.
          Then the value of the parameter is given by the calculation
         as shown in the third column. }
\begin{tabular}{ccl}
 parameter & measurement & calculation \\ \tableline
 $ \Rp \Delta $
&
 $ \left( \alpha_l(t), A \right) $
&
 $ \frac{1}{2}  \left[ \left( \alpha_l(t), A \right)
 - 2 \Rp y_l - 2 \Rp \BD \right] $
\\ \tableline
 $ \Rp( x_l - \bar{x}_l ) $
&
 \begin{tabular}{c}
  \\
  $ \left( \beta_l(t), A \right) $
 \end{tabular}
&
 \begin{tabular}{l}
  $ 2 \Rp \BD - 2 \Rp \Delta $ \\
  $ \frac{1}{2} \left[ - \left( \beta_l(t), A \right)
  + 4 \Rp \epsilon_0 - 2 \Rp y_l \right] $
 \end{tabular}
\\ \tableline
 $ \displaystyle{
      \left( 1 - \frac{ \Rp \bar{A}_0 }{ \Rp A_0 } \right)
   } $
& & From $ 2 \Rp( \epsilon - ( \epsilon_0 - \Delta ) ) $
\\ \tableline
 $ \displaystyle{ \left( 1 - \frac{ \Rp \bar{A}_2 }{ \Rp A_2 }
                  \right)
                } $
& &
 $ \displaystyle{ \Rp \left(
                       \frac{ 2 \sqrt{2} \epsilon^\prime }{ \omega }
                        e^{ -i( \delta_2 - \delta_0 ) }
                      \right)
                    + \left(
                       1 - \frac{ \Rp \bar{A}_0 }{ \Rp A_0 }
                      \right)
                } $
\end{tabular}
\label{ tab:abexp }
\end{table}
%
%
%%%%%%%%%%%%%%%%%%%%%%%%%%%%%%%%%%%%%%%%%%%%%%%%%%%%%%%%%%%%%%
%                                                            %
%                          Table 4                           %
%                                                            %
%%%%%%%%%%%%%%%%%%%%%%%%%%%%%%%%%%%%%%%%%%%%%%%%%%%%%%%%%%%%%%
%
%%
\begin{table}
\caption{ The coefficients $ A-H $ in the time dependent asymmetries,
         functional form of which is of the type $ \Lambda $
         ( see Eq.$ \left( \ref{ eqn:Lambda } \right) $ ).}
%
%\begin{center}
%
\begin{tabular}{ccccc}
 & $ A $ & $ B $ & $ C $ & $ D $ \\
\tableline
 \( \delta_l \)
 &
$\begin{array}{l}
 \Rp( \epsilon_0 - y_l ) \\
 - \Rp \BD
\end{array}$
 &
$\begin{array}{l}
 \Rp(\epsilon_0 - y_l ) \\
 + \Rp \BD
\end{array}$
 & \( 1 - 2 \Rp \epsilon_0 \)
 &
$\begin{array}{l}
 - 2 \Ip \Delta \\
 - \Ip( x_l - \bar{x}_l )
\end{array}$
\\ \tableline
 \( \bar{\delta}_l \)
&
$\begin{array}{l}
 \Rp( \epsilon_0 - y_l ) \\
 - \Rp \BD
\end{array}$
 &
$\begin{array}{l}
 \Rp( \epsilon_0 - y_l ) \\
 + \Rp \BD
\end{array}$
 & \( -1 - 2 \Rp \epsilon_0 \)
 &
$\begin{array}{l}
 -2 \Ip \Delta \\
 + \Ip( x_l - \bar{x}_l )
\end{array}$
\\ \tableline
 \( \gamma_l \) & \( \Rp \epsilon_0 + \Rp \Delta \)
& \( \Rp \epsilon_0 - \Rp \Delta \)
& \( - 1 - 2 \Rp( \epsilon_0 - y_l ) \)
& \( 2( \Ip \Delta + \Ip x_l ) \)
\\ \tableline
 \( \bar{\gamma}_l \) & \( \Rp \epsilon_0 + \Rp \Delta \)
& \( \Rp \epsilon_0 - \Rp \Delta \)
& \( 1 - 2 \Rp( \epsilon_0 - y_l ) \)
& \( 2 ( \Ip \Delta + \Ip \bar{x}_l ) \)
\\
\end{tabular}
\label{ tab:lambdacoeff }
%
%\end{center}
%
\end{table}
%
%%%%%%%%%%%%%%%%%%%%%%%%%%%%%%%%%%%%%%%%%%%%%%%%%%%%%%%%%%
%                                                        %
%                    Table 4 (continued)                 %
%                                                        %
%%%%%%%%%%%%%%%%%%%%%%%%%%%%%%%%%%%%%%%%%%%%%%%%%%%%%%%%%%
%
%
\begin{table}
%
%\begin{center}
%
\begin{tabular}{ccccc}
 & $ E $ & $ F $ & $ G $ & $ H $ \\ \tableline
 \( \delta_l \) &
$\begin{array}{c}
 2 \Rp \Delta + \Rp x_l \\
 + \Rp \bar{x}_l + 2 \Rp \epsilon_0
\end{array}$
&
$\begin{array}{c}
 -[ 2 \Rp \Delta + \Rp x_l \\
 + \Rp \bar{x}_l ] + 2 \Rp \epsilon_0
\end{array}$
& \( -4 \Rp( \epsilon_0 - y_l ) \)
&
$\begin{array}{ll}
 & 4 \Ip \Delta \\
 & + 2 \Ip( x_l + \bar{x}_l )
\end{array}$
\\ \tableline
 \( \bar{\delta}_l \) &
$\begin{array}{c}
 2 \Rp \Delta - \Rp x_l \\
 - \Rp \bar{x}_l + 2 \Rp \epsilon_0
\end{array}$
&
$\begin{array}{c}
 -[ 2 \Rp \Delta - \Rp x_l \\
 - \Rp \bar{x}_l ] + 2 \Rp \epsilon_0
\end{array}$
& \( -4 \Rp( \epsilon_0 - y_l ) \)
&
$\begin{array}{ll}
 & 4 \Ip \Delta \\
 & + 2 \Ip( x_l + \bar{x}_l )
\end{array}$
\\ \tableline
 \( \gamma_l \)
&
$\begin{array}{c}
 2[ \Rp( \epsilon_0 - y_l ) \\
 - \Rp \Delta - \Rp x_l ]
\end{array}$
&
$\begin{array}{c}
 2[ \Rp( \epsilon_0 - y_l ) \\
 + \Rp \Delta + \Rp x_l ]
\end{array}$
& \( -4 \Rp \epsilon_0 \) & \( -4 \Ip \Delta \)
\\ \tableline
 \( \bar{\gamma}_l \)
&
$\begin{array}{c}
 2[ \Rp( \epsilon_0 - y_l ) \\
 - \Rp \Delta + \Rp \bar{x}_l ]
\end{array}$
&
$\begin{array}{c}
 2[ \Rp( \epsilon_0 - y_l ) \\
 + \Rp \Delta - \Rp \bar{x}_l ]
\end{array}$
& \( -4 \Rp \epsilon_0 \) & \( -4 \Ip \Delta \)
\\
\end{tabular}
%
%\end{center}
%
\end{table}
%
%
%%%%%%%%%%%%%%%%%%%%%%%%%%%%%%%%%%%%%%%%%%%%%%%%%%%%%%%%%%%%
%                                                          %
%                         Table 5                          %
%                                                          %
%%%%%%%%%%%%%%%%%%%%%%%%%%%%%%%%%%%%%%%%%%%%%%%%%%%%%%%%%%%%
%
%
\begin{table}
\caption{ The determination of the values of parameters
         from $ \delta_l(t),\bar{\delta}_l(t),\gamma_l(t) $
         and $ \bar{\gamma}_l(t) $ in the kaon tagging experiments.
          The meaning of $ ( \delta_l, E ) $,
         for example, is the same as
         in Table \ref{ tab:abexp }. }
\begin{tabular}{ccl}
 parameter & measurement & calculation
\\ \tableline
 $ \Ip x_l + \Ip \bar{x}_l $ & & $ 2 \Ip \BD - 2 \Ip \Delta $
\\ \tableline
 $ \Rp( x_l + \bar{x}_l ) $
&
 \begin{tabular}{c}
  $ \left( \delta_l(t), E \right) $ \\
  $ \left( \bar{ \delta }_l(t), E \right) $
 \end{tabular}
&
 \begin{tabular}{l}
  $ \left( \delta_l(t), E \right)
    - 2 \Rp \Delta - 2 \Rp \epsilon_0 $ \\
  $ 2 \Rp \Delta + 2 \Rp \epsilon_0 - \left( \bar{ \delta }_l(t),
    E \right) $
 \end{tabular}
\\ \tableline
 $ \Rp x_l $
&
 \begin{tabular}{c}
   \\
  $ \left( \gamma_l(t), E \right) $
 \end{tabular}
&
 \begin{tabular}{l}
  $ \frac{1}{2} \left[ \Rp( x_l + \bar{x}_l )
      + \Rp( x_l - \bar{x}_l ) \right] $ \\
  $ \Rp( \epsilon_0 - y_l ) - \Rp \Delta
      - \left( \gamma_l(t), E \right) $
 \end{tabular}
\\ \tableline
 $ \Rp \bar{x}_l $
&
 \begin{tabular}{c}
   \\
  $ \left( \bar{ \gamma }_l(t), E \right) $
 \end{tabular}
&
 \begin{tabular}{l}
  $ \frac{1}{2} \left[ \Rp( x_l + \bar{x}_l )
       - \Rp( x_l - \bar{x}_l )
    \right] $ \\
  $ \left( \bar{ \gamma }_l(t), E \right) - \Rp( \epsilon_0 - y_l )
  + \Rp \Delta $
 \end{tabular}
\\ \tableline
 $ \Ip x_l - \Ip \bar{x}_l $
&
 \begin{tabular}{c}
  $ \left( \delta_l(t), D \right) $ \\
  $ \left( \bar{ \delta }_l(t), D \right) $
 \end{tabular}
&
 \begin{tabular}{l}
  $ -2 \Ip \Delta - \left( \delta_l(t), D \right) $ \\
  $  2 \Ip \Delta + \left( \bar{ \delta }_l(t), D \right) $
 \end{tabular}
\\ \tableline
 $ \Ip x_l $
&
 \begin{tabular}{c}
   \\
  $ \left( \gamma_l(t), D \right) $
 \end{tabular}
&
 \begin{tabular}{l}
  $ \frac{1}{2} \left[ \Ip( x_l + \bar{x}_l )
       + \Ip( x_l - \bar{x}_l ) \right]
  $ \\
  $ \left( \gamma_l(t), D \right) - \Ip \Delta $
 \end{tabular}
\\ \tableline
 $ \Ip \bar{x}_l $
&
 \begin{tabular}{c}
   \\
  $ \left( \bar{ \gamma }_l(t), D \right) $
 \end{tabular}
&
 \begin{tabular}{l}
  $ \frac{1}{2} \left[ \Ip( x_l + \bar{x}_l )
       - \Ip( x_l - \bar{x}_l )    \right] $ \\
  $ \left( \bar{ \gamma }_l(t), D \right)- \Ip \Delta $
 \end{tabular}
\\
\end{tabular}
\label{ tab:lambdaexp }
\end{table}

\begin{references}
\bibitem{Frascati}
  Plans to build the $ \phi $ factory are also being studied
 at KEK \cite{KEK} and UCLA ( see the proceeding cited
 in \cite{LEAR} ).
\end{references}
\end{document}